\documentclass{aa} 
\usepackage[authoryear]{natbib}
\bibpunct{(}{)}{;}{a}{}{,}
\usepackage{amsmath}
\usepackage{graphicx}
\usepackage{lscape}
\usepackage{soul}
\usepackage{ulem}
\usepackage{color}

\begin{document}

\titlerunning{Equalizing SPH using $sinc$ kernels} 
\authorrunning{Garc\'{i}a-Senz, Cabez\'on,  Escart\'in \& Ebinger}

\title{Equalizing resolution in smoothed-particle hydrodynamics calculations using self-adaptive sinc kernels}

\author{Domingo Garc\'{i}a-Senz\inst{1}, 
       Rub\'en M. Cabez\'on\inst{2}, 
        Jos\'e A. Escart\'{i}n\inst{1},
          \and
        Kevin Ebinger\inst{2}
       }
\institute
{Dept. de F\'{i}sica i Enginyeria Nuclear, Universitat Polit\`ecnica de Catalunya. Compte d'Urgell 187, 08036 Barcelona (Spain) and Institut d'Estudis Espacials de Catalunya. Gran Capit\`a 2-4, 08034 Barcelona, Spain \\
 \and
Departement Physik. Universit\"at Basel. Klingelbergstrasse 82, 4056 Basel, Switzerland \\}

\date{}

\abstract
{The smoothed-particle hydrodynamics (SPH) technique is a numerical method for solving gas-dynamical problems. It has been applied to simulate the evolution of a wide variety of astrophysical systems. The method has a second-order accuracy, with a resolution that is usually much higher in the compressed regions than in the diluted zones of the fluid.   
}
{We propose and check a method to balance and equalize the resolution of SPH between high- and low-density regions. This method relies on the versatility of a family of interpolators called $sinc$ kernels, which allows increasing the interpolation quality by varying only a single parameter (the exponent of the $sinc$ function).  
}
{The proposed method was checked and validated through a number of numerical tests, from standard one-dimensional Riemann problems in shock tubes, to multidimensional simulations of explosions, hydrodynamic instabilities, and the collapse of a Sun-like polytrope.    
}
{The analysis of the hydrodynamical simulations suggests that the scheme devised to  equalize the accuracy improves the treatment of the post-shock regions and, in general, of the rarefacted zones of fluids while causing no harm to the growth of hydrodynamic instabilities. The method is robust and easy to implement with a low computational overload. It conserves mass, energy, and momentum and reduces to the standard SPH scheme in regions of the fluid that have smooth density gradients.  
}
{}   


\maketitle

\section{Introduction}

The hydrodynamical method known as smoothed-particle hydrodynamics (SPH) is a gridless Lagrangian approach to continuum mechanics  devised by \citet{gm77}~and \citet{lucy77}. A key ingredient of the SPH technique is the nature of an interpolating function called the kernel, which is used to estimate the value of different physical magnitudes. Because of the nature of the SPH interpolations, the gradient of any magnitude can be calculated by directly taking the gradient of the kernel, which is an analytically differentiable function. This provides an easy and efficient way of obtaining gradients. In that way it is easy to write the Euler equations of fluid mechanics in terms of the kernel and its derivatives \citep{monaghan92, monaghan05}.
Despite its success, SPH still has several weak points, which have recently caused a number of improvements of the technique \citep{rosswog2014,saitoh2013, garciasenz12, cabezon12, valdarnini12, dehnen12, springel2010a}. One of the shortcomings of SPH is that the accuracy in the density evaluation is  different in all fluid
regions. The resolution in low-density regions is typically poorer than in the high-density regions. In astrophysics, the regions close to the surface of self-gravitating bodies usually  have
a lower resolution than the interior. Another source of inaccuracy are fluid discontinuities, such as shock waves or sharp boundaries. In this case, the difficulty comes from the inefficacy of the interpolations to keep track of phenomena with a length-scale lower than the characteristic smoothing length $h$. 

It is well known that the standard formulation of SPH is second-order accurate in $h$. An interpolating function $W = (K/h)~f(v)$~is defined so that the averaged value of density at any point of the system is estimated (in 1D) as   

\begin{equation}
<\rho(x_0)>=\rho_0- \frac{1}{2}\frac{d^2\rho}{dx^2}~h^2~K~\int_{-\infty}^{+\infty}v^2~f(v)~dv + \phi(h^4)+ ....
\label{2order}
,\end{equation}

\noindent
where $K$~is a normalization constant and $v=\vert x-x_0\vert/h$, where $h$~is a scaling parameter called the smoothing-length. The magnitude  $<\rho(x_0)>$ is the SPH estimation of density at the fluid coordinate $x_0$, and $\rho_0$ is the value of the density at that point. The smoothing length $h$ is usually taken as the local resolution. Therefore, the lead dependence of the error in evaluating the density is proportional to $h^2$. In most applications the value of the smoothing-length is directly linked to the density via mass conservation, $h\propto\rho^{-1/d}$, where $d$ is the dimension of the space, resulting in a self-consistent Lagrangian description of the dynamics \citep{springel02}. However, setting the smoothing length in that way  is detrimental to the low-density regions, where $h$ becomes large. On another note, the error term in Eq. (\ref{2order}) has a dependence on the second derivative of the smoothed function itself, strictly vanishing only for linear functions. As a consequence, one can expect relevant differences in accuracy along the fluid wherever the second derivative of the function does not vanish. 

To this extent, we can use the shape of the kernel as a parameter to gain additional control over the error in Eq.~(\ref{2order}) so that the interpolations in density are made with approximately the same relative accuracy in any point of the system. An easy way of achieving this is using the $sinc$ kernels.

In this paper we introduce a novel method to balance the resolution in all regions of the fluid. We take advantage of the features of the one-parametric family of kernels introduced by \cite{cabezon08} to set an additional variable (the exponent $n$ of the $sinc$ function, given by Eq.~2) which, jointly with $h$,~controls the local resolution. We found a mechanism to increase the accuracy of interpolations by changing $n$ when a strong density gradient is detected. In fluid regions with low or moderate gradients, the scheme reduces to the standard \citep{monaghan05}, where the resolution is basically set by the smoothing length.

The paper is organized as follows: In Sect. 2, the main features of the proposed family of kernels are reviewed and compared with other existing interpolators. The numerical scheme used to equalize the resolution is explained in Sect. 3, where we also provide some insights into the ability of the algorithm to handle sharp one-dimensional density profiles. The basic Euler equations, written in the SPH formalism and incorporating the adaptive $sinc$ kernels, are described in Sect. 4 and in Appendix A. These sections also include the treatment of grad-h and grad-n terms, as well as details of the practical implementation of the algorithm. In Sect. 5, we check the hydrocode with a variety of standard tests in one, two, and three dimensions. These tests intend to cover different physical processes, such as shocks, instabilities, or self-gravitating bodies. Finally, we outline the main conclusions of our work and comments on the shortcomings of the developed scheme as well as on future lines of improvement in Sect. 6.

\section{Main features of the $sinc$ family of kernels}
\label{section2}
The $sinc$ family of compact supported kernels was first introduced by \citet{cabezon08} and \citet{cabezon12}  (who called them
{\sl harmonic kernels}) as a way to make the SPH technique more flexible. The ultimate goal was to gather in a unique function the more relevant features of some of the most often used
interpolators. The $sinc$ family of kernels $W_n^s$ is defined as

\begin{equation}
W_n^s(v,h,n)=\frac{B_n}{h^d} S_n (\frac{\pi}{2}v)\qquad 0\leq v\leq 2, 
\label{sinc}
\end{equation}

\noindent  where $S_n(.)=sinc^n(.)$, and where $n$ is the index of the kernel and $B_n$ a normalization constant. The function $sinc (\frac{\pi}{2}v)=\left[\frac{sin(\frac{\pi}{2}~v)}{(\frac{\pi}{2}v)}\right]$ is a widely known function used in signal analysis and spectral theory. The main advantage of this kernel is that it is able to mimic the behavior of some of the most popular kernels, such as the cubic and quintic splines \citep{cabezon08}. By a careful choice of the index $n$, it can even approach several of the so-called Wendland kernels \citep{wendland95}, recently discussed in \citet{dehnen12}, which behave optimally in avoiding pairing instability. In Fig.~\ref{figure1} we present the profiles of several $sinc$, $S_n$ , and Wendland $\psi_{l:k}$ kernels with the independent variable rescaled to the range $[0,1]$. The profiles of the high order $\psi_{3:1}$, $\psi_{4:2}$ , and $\psi_{5:3}$ Wendland interpolators can be approached  by $S_{3.97}$, $S_{5.12}$ , and $S_{6.41}$ respectively, whereas several members of the  spline family can be approached by $S_{3.07}$ (cubic spline), $S_{3.97}$ (quartic spline), and $S_{4.89}$ (quintic spline). According to \citet{dehnen12}, a necessary condition to escape the pairing instability is that the Fourier transform, $\mathcal F_3 [W(r)](\kappa)=4\pi\kappa^{-1}\int_0^\infty~\sin(\kappa r)~W(r)~rdr$, (in 3D, where $\kappa$ is the wavenumber)~of the kernel is always definite positive. In Fig.~\ref{figure2} we show $\mathcal F_3 (S_n)$ for kernel indexes $n=3, 6, 10$. For $n= 3$ (similar to the cubic spline) the Fourier transform becomes negative at relatively low wavelengths, limiting the number of neighbors in the summations in practical applications. Nevertheless, for $n=6$ the dynamical range has been considerably extended and is now similar to that of the quintic spline. For  $n= 10$ the Fourier transform becomes marginally negative at long wavelengths, thus showing a good endurance in front of the pairing.

The relationship between the index of the kernel, $n$,~and the maximum number of neighbors (to elude particle clustering), $\mathcal N$,~can be made more explicit by applying the empirical rule described in \citet{price2012}. Many computer simulations have shown that particle clustering is avoided when the normalized interparticle separation in ordered lattices becomes larger than a critical value,  $r_{ij}/h\ge \eta$, with  $\eta\simeq 1.2$~for the cubic spline~\citep{price2012}. A value $\eta= 1.2$ corresponds to $\mathcal N_{max}=18,~ 57$ neighbors in 2D and 3D, respectively, so that for $\mathcal N > \mathcal N_{max}$ there is an increasing chance for particle clustering. The second derivative of the $sinc,~ n=3$ (cubic-spline-like kernel), vanishes at $\eta'=r_{ij}/h=1.5,$ which is a factor $f=5/4$ larger than $\eta=1.2$. To find $\mathcal N_{max}$ for any $n,$ we first estimated the point $\eta'$ where the second derivative of $S_n$~vanishes as a function of the kernel exponent $n$ and calculated the value of $\eta (n)= \eta' f^{-1}$. That value of $\eta (n)$ was then mapped to $\mathcal N_{max}$ (i.e., the maximum amount of neighbors that is still resistant to particle clustering). In Fig.~\ref{figure3} we depict the profile of the maximum number of neighbors as a function of the  kernel index $n$ in two and three dimensions. For $\mathcal N\simeq 60, 100,~200$ neighbors in 3D it is advisable to take $n\geq 3, 4,~\mathrm{and}~7,$ respectively, in Eq.~(\ref{sinc}) to avoid particle pairing.

The minimum number of neighbors $\mathcal N_{min}$ is also constrained by the approach of SPH integrals as finite summations. The precise value of $\mathcal N_{min}$ is difficult to estimate because
it ultimately depends on the specific physical problem. A simple, albeit qualitative, way to set $\mathcal N_{min}$ is to numerically calculate the value of the density in a regular $bcc$ lattice as a function of the kernel index $n$ and the number of neighbors, and measure of the relative deviation of the density from the theoretical value. To do this, we built 2D and 3D regular square lattices of unit square and placed $N$ particles at the nodes of the grid. By giving a mass $m=1/N$ to each particle, this should result in a uniform density $\rho=1$. As a practical criterion for selecting $\mathcal N_{min}$, we calculated the density, $\rho_{SPH}$, for many pairs $(n,\mathcal N)$ as well as the value of $\sigma=(\vert\rho_{SPH}-1\vert)$. These combinations with $\sigma \geq 10^{-3}$ were then considered not a good enough realization of density, and the critical limiting values with $\sigma\simeq 10^{-3}$ were stored. The results of this study is summarized in the dotted blue lines shown in Fig. \ref{figure3}. The most striking feature of these lines is that they have a minimum for kernel indexes $n = 4-6,$ which indicates that these $sinc$ interpolators are the best choice to carry out SPH calculations. These results support the current feeling that using the quarter or the quintic spline enhances the convergence of interpolations. It is worth noting that when the pairing and the convergence criteria are combined, an optimal domain in the plane $(n, \mathcal N) $ appears, which, for a given kernel index $n$, restricts the number of neighbors to $\mathcal N_{min}\leq\mathcal N\le \mathcal N_{max}$. It is equally remarkable that the pairing and the convergence lines intersect at $n\simeq 3$ in 2D and 3D, meaning that a large body of  SPH calculations carried out so far with the cubic-spline kernel have probably been bordering on an unpredictable and dangerous zone. It should be kept in mind, however, that the path of these lines in the $(n, \mathcal N)$ diagram is merely qualitative, and our advice is not to proceed too far into the forbidden regions in practical applications.

The implementation of the $W_n^s(v,h,n)$ family of kernels adds more flexibility to SPH because one can, for example, take a different index $n$ to handle the artificial viscosity terms in the momentum and energy equations or in the heat conduction equation, without changing the number of neighbors of the particle. Another virtue of the $W_n^s(v,h,n)$ family is that by a careful choice of kernel exponents it allows equalizing the interpolation accuracy along the system; this last point is the subject of the present study.

To increase the computational speed, it is recommended to store the value of ${\mathrm sinc\left(\frac{\pi}{2} v\right)}$ and its derivative as a function of \mbox{$v,\ (0\le v\le 2)$} in a table and use a linear Taylor expansion to calculate the value of $\mathrm sinc$ and other variables of interest (see Appendix A). This allows a fast computation of Eq.~(\ref{sinc}) and its derivative after the index $n$ is chosen. The value of the normalization constant $B_n$ for $2\le n\le 12$ can be obtained from the following fitting functions:

\begin{equation}
B_n=
\left\{\begin{array}{rclcc}
b_0+b_1 n^{1/2}+b_2 n+b_3 n^{-1/2} & \qquad\mathrm{1D} \\
b_0+b_1 n+b_2 n^{-1}+b_3 n^{-2} & \qquad\mathrm{2D} \\
b_0+b_1 n^{1/2}+b_2 n+b_3 n^{3/2} & \qquad\mathrm{3D},
\end{array}
\right.
\label{normalization}
\end{equation}

\noindent where the values of coefficients $b_0, b_1, b_2,b_3$ as a function of the dimensionality are provided in Table 1. These fitting functions are fast to compute and precise up to the fifth significant figure. 

\section[]{Using the $sinc$ kernels to enhance interpolations}

According to Eq. (\ref{2order}), the leading error ${\mathcal E}$ in estimating density
is
\begin{equation}
\vert\mathcal E\vert=h^2~\frac{1}{2}~\vert\nabla^2\rho\vert{\mathcal I(n)} 
\label{error}
,\end{equation}

\noindent
where $n$ and $h$ are the kernel index and the smoothing length. In the standard Lagrangian formulation of SPH \citep{monaghan05} the smoothing length is constrained so that the neighboring mass of a given particle always remains constant. This fact leaves the kernel index $n$ as the only free parameter to control the error size given by Eq.~(\ref{error}). In general, increasing  the value of $n$ decreases the error, but one cannot increase $n$ arbitrarily without generating
too much numerical noise.

It is easy to estimate the relative accuracy achieved with several exponents $n$ using Eq.~(\ref{error}). First we write the error term as

\begin{equation}
{\frac{2\mathcal E}{\vert\nabla^2\rho\vert}= h^2 \mathcal I_n}
\label{accuraccy1}
,\end{equation}

with 

\begin{equation}
\mathcal I_n= A\pi B_n \int_0^{\infty} v^p sinc^n(\frac{\pi}{2}~v) dv
\label{int_n}
,\end{equation}
 
\noindent
with $A= 2, 4$, and $p=3, 4$ in 2D and 3D, respectively, and $B_n$ is the normalization constant given by Eq.~(\ref{normalization}). Setting arbitrarily $\frac{2\mathcal E}{\vert\nabla^2\rho\vert}=1$ for $n=3$, $\mathcal N=57$ neighbors in 3D and for $n=3$, $\mathcal N=18$ in 2D (i.e., we choose the normalization for the error), and taking into account that $h\propto\mathcal N ^{1/d}$ , we write 

\begin{equation}
\frac{2\mathcal E}{\vert\nabla^2\rho\vert}=
\left\{\begin{array}{rclcc}
\left(\frac{\mathcal N}{18}\right)\frac{\mathcal I_n}{\mathcal I_3}\qquad\qquad \mathrm{2D}\\

\left(\frac{\mathcal N}{57}\right)^{\frac{2}{3}}\frac{\mathcal I_n}{\mathcal I_3}\qquad\quad\mathrm{3D}.
 \\
\end{array}
\right.
\label{errors}
\end{equation}

For a given kernel index $n$ the integral $\mathcal I_n$ is calculated numerically and the value of the magnitude $2\mathcal E/\vert\nabla^2\rho\vert$ as a function of pairs $(n,\mathcal N)$ is shown in Fig.~\ref{figure3}. Although $2\mathcal E/\vert\nabla^2\rho\vert$ is not directly the interpolation error, it does give an estimate of it. Therefore, for the sake of simplicity, we refer to this magnitude as the
error in the following. From this figure we see, for example, that a similar accuracy is achieved by different pairs $(n, \mathcal N)$. For example, the $S_3$~$sinc$ kernel using $\mathcal N=57$ neighbors in 3D has a similar leading error term as $S_5$ and $\mathcal N=100$ or $S_{10}$ and $\mathcal N=250$. On the whole, computations are fast for low $\mathcal N,$ but the results are more sensitive to numerical noise, while the opposite is true for high $\mathcal N$. A conservative option is to work with a moderate number of neighbors and variable kernel indexes, leaving  high kernel exponents to handle only regions with steep gradients. As a default, we took $n=5$ and $\mathcal N\simeq 50,~100$ neighbors in the $2D$ and $3D$ numerical experiments described in Sect.~5, although other combinations are also feasible. A similar number of neighbors and a high B-spline kernel, quartic ($\simeq S_4$) or quintic ($\simeq S_5$), was suggested by \cite{valdarnini12} as an optimal choice to improve the convergence of hydrodynamic simulations.

However, it is not straightforward to use Eq.~(\ref{error}) to control the error $\mathcal E$ via changing the kernel index $n$ in $\mathcal I(n)$,  owing to the dependence of the expression on the second derivative of the function. Instead, we introduced an estimator parameter $\lambda$, such that $\lambda=1$ when the system behaves linearly, but $\lambda > 1$ in regions where the fluid departs from linearity. The local value of $\lambda$ was then used to set the exponent $n$ of the kernel that reduces the error in the density estimation. A similar strategy was introduced by \citet{sigalotti06}~to 
select the value of the smoothing parameter $h$. We comment on the similarities and differences between our proposal and that of Sigalotti and coworkers in the concluding section.

For each particle $a$ an estimator $\lambda_a$ is defined,

\begin{equation}
\lambda_a=
\left\{\begin{array}{rclcc}
\left(\frac{\bar\rho_a}{\rho_a}\right) & \mathrm{for}~ & \bar\rho_a\geq\rho_a, \\
\left(\frac{\rho_a}{\bar\rho_a}\right) & \mathrm{for}~ & \bar\rho_a <\rho_a, 
\end{array}
\right.
\label{estimator}
\end{equation}

\noindent
where $\ln\bar\rho_a=\frac{1}{\mathcal N_a}\sum_{b=1}^{\mathcal N_a} \ln\rho_b$ , where $\rho_b$ is a density estimate calculated with a kernel index $n$, and $\mathcal N_a$ is the number of neighbors of the particle.
When the values of $\lambda_a$ are known, a new kernel index $n_a$ is assigned to each mass point  according to 

\begin{equation}
n_a=n_{0}+\Delta n\cdot f(\xi_a),\qquad\mathrm{with}\qquad \xi_a=\frac{(\lambda_a-1)}{\lambda_c}~\geq 0\,,
\label{indexes}
\end{equation}

\noindent  where $n_0$~is a constant  $baseline$ value of the kernel index set at the beginning of the calculation, $\Delta n$~is the highest allowed jump above the baseline value, and  $\lambda_c$ is a scaling parameter ($\lambda_c\simeq 0.5$). At each model a trial value of $h$ and $n$ are picked and the  density is computed. These trial values are iteratively refined, with the scheme explained in Appendix A, until the constraints  
on $h$ ($h^d\rho=$ constant) and $n$ (Eq.~\ref{indexes}) are fulfilled.  $n=n_0$ can usually be taken
everywhere for the first model, but it departs from $n_0$ in fluid regions with a nonlinear behavior. Below we refer to the adaptive $sinc$ kernel indexes as $n(x)$ in 1D numerical experiments with static configurations, $n(x,t)$ for time-dependent 1D hydrodynamic simulations, and $n({\bf r}, t)$ in more than one dimension.

The function $f(\xi)$ must fulfill at least two limiting conditions: 

\begin{equation}
\left\{\begin{array}{rclcc}
\lim_{\xi\to 0} f(\xi)\rightarrow 0, \\
\lim_{\xi\to \infty} f(\xi)\rightarrow 1. 
\end{array}
\right.
\label{limits}
\end{equation}

This ensures that the kernel index remains close to its baseline value, $n_0$, in regions where $\bar\rho_a\simeq\rho_a,$ but becomes $n\simeq n_0+\Delta n$ in regions with a clearly nonlinear behavior. 

A suitable function $f(\xi)$ used in this work is

\begin{equation}
f(\xi)=1-\frac{2}{\exp(\xi)+\exp(-\xi)}
\label{ffunction}
.\end{equation}

This function has the interesting property that $df(\xi)/d\xi)\simeq 0$ at low $\xi$, making it insensitive to numerical noise. Nevertheless, the function becomes steep at moderate $\xi$ while flattering again at $\xi >> 1$. 
    
An important feature of this scheme is that it is compatible with the Lagrangian derivation of the SPH equations. In other words, the gradient of the kernel index can be incorporated into the Euler equations in the same way as the gradient of the smoothing length is taken into account in the standard SPH \citep{springel02}.  The reason is that the estimator $\lambda_a$ defined by Eq. (\ref{estimator})~admits an explicit derivative with respect to $\rho_a$,

\begin{equation}
\frac{\partial\lambda_a}{\partial\rho_a}=
\left\{\begin{array}{rclcc}
\frac{\lambda_a}{\rho_a}\left(\frac{1}{\mathcal N_a}-1\right) & \mathrm{for}~ & \bar\rho_a\geq\rho_a, \\
\frac{\lambda_a}{\rho_a}\left(1-\frac{1}{\mathcal N_a}\right) & \mathrm{for}~ & \bar\rho_a <\rho_a. 
\end{array}
\right.
\label{Derivestimator}
\end{equation}

Using Eqs.~(\ref{indexes}), (\ref{ffunction}), and (\ref{Derivestimator}), is straightforward to compute $\frac{\partial n}{\partial\rho}$, which is needed to correct the Euler equations from the new grad-n terms (see Sect.~4). The constraint in $n$ set by Eq.(\ref{indexes}) is, however, of a different kind than that arising from $h^d\rho=$ constant (used to set the value of the smoothing length at each time step). While the latter is a real physical constraint and
a direct consequence of mass conservation, the former arises from a mathematic consideration of linearity in a local fluid
region. 

\subsection{Fitting sharp 1D density profiles} 

In this section we consider the ability of the proposed scheme to reproduce the 1D density profile of some mass distributions  that often appear in hydrodynamics calculations.  These profiles are referred to as {\it mountain}, {\it valley}, {\it wall,}~and {\it cliff}~ (see Fig.\ref{figure4}). Their mathematical expressions are
\begin{equation}
\rho(x)=\rho_0+\Delta\rho~e^{-(\frac{x-x_0}{\delta})^2},\qquad\mathrm {(mountain)} 
\label{mountain}
\end{equation}
 
\begin{equation}
\rho(x)=\rho_0-\Delta\rho~e^{-(\frac{x-x_0}{\delta})^2},\qquad\mathrm {(valley)}
\label{valley}
\end{equation}

\begin{equation}
\rho(x)=\rho_0+\Delta\rho~\frac{e^{(\frac{x-x_0}{\delta})}-e^{-(\frac{x-x_0}{\delta}})}{e^{(\frac{x-x_0}{\delta})}+e^{-(\frac{x-x_0}{\delta}})},\qquad\mathrm {(wall)}
\label{wall}
\end{equation}

\begin{equation}
\rho(x)=
\left\{\begin{array}{rclcc}
\rho_0 &  \mathrm{for}~ & x<x_0 \\
\rho_0~e^{-(\frac{x-x_0}{
\delta})}& \mathrm{otherwise},&\qquad\mathrm {(cliff)} 
\end{array}
\right.
\label{cliff}
\end{equation}

\noindent 
where $\delta$ is the characteristic width of the function. The convolution of this curve with the kernel provides the SPH density
values. The density is calculated with the standard SPH summation 

\begin{equation}
\rho_a=\sum_{b=1}^{\mathcal N_a} m_b W^s_{ab}(x_a,x_b,h_a,n_a)
\label{dens}
.\end{equation}

The parameter values in expressions (\ref{mountain}-\ref{cliff}) are shown in Table~2, and $\delta < h$, $\mathcal N_a=100,\forall a$. The results of the calculations are depicted in Fig. 4. These profiles are idealized mathematical curves mimicking physical situations of considerable interest in gas dynamics. A mountain-like profile with density contrast of four can appear in regions with a strong shock moving through a perfect gas with adiabatic index $\gamma=5/3$. The inverted Gaussian (valley-like) structures may appear during the propagation rarefaction waves. Wall-like structures can be found at fluid regions that contact rigid boundaries. Self-gravitating bodies usually end in rarefied atmospheres with steep (cliff-like) density gradients. In all the curves shown in Fig.\ref{figure4}, the characteristic width in the steepest regions is lower than the smoothing length. Thus we expect problems in the SPH approach for particles located in the neighborhood of the discontinuities.

The analysis of  these idealized curves unambiguously indicates an enhancement of the numerical fitting when adaptive kernel indexes are used. The improvement is especially good in the low-density regions that host steep gradients (see, for example, the upper
right panel in Fig.\ref{figure4} that shows an inverted Gaussian). These rarefacted regions are precisely the regions where the standard SPH gives the poorest results because the smoothing length $h$ becomes longer to satisfy the constraint $\rho h^d=$ constant. 

The comparison between the {\it mountain} and {\it valley}-like profiles suggests that the effect of equalizing the error is not symmetric. In the first case the highest value of $n$ ($\simeq 8)$ is achieved not at the peak of the Gaussian, but at the base of the profile where the curve becomes flat, while in the second case $n\simeq 8$ is taken just at the bottom of the inverted Gaussian. In both cases the index $n$ clearly increases in regions where the second derivative does not vanish. The $\it wall$-like case is similar to that of the $\it mountain,$ but with a plateau to the right of the profile. Again it can be seen as $n$ peaks at the base of the $\it wall$ where the second derivative is larger, maximizing the differences among the profiles calculated with $n=3$, $n=5$ and $n(x)$. Finally, the case of the $\it cliff$, depicted in the bottom-right panel of Fig.~\ref{figure4}, is  particular because the particle sample near the base of the $\it cliff$ is usually sparse, if not void (especially in 3D applications). Even though the true profile is not reproduced by any of the calculations, we see that the better fit is achieved using variable exponents. This suggests that using adaptive $sinc$ kernels might be of great interest to simulate phenomena in the envelope of self-gravitating bodies, as long as a sufficient sample of particles is available.         

\section[]{Hydrodynamic equations}

The numerical scheme described above was validated through several
hydrodynamic tests and compared with the results obtained from
keeping  the kernel exponent unchanged. We used the standard SPH written in the Lagrangian formulation as described in \citet{monaghan05}, \citet{rosswog09}, \citet{springel2010b}, and \citet{price2012}.

\subsection{Euler equations with grad-h and grad-n terms}

Small changes of the standard SPH scheme are necessary to incorporate the $sinc$ family of kernels with adaptive indexes. The Euler equations do not change:    

\begin{equation}
\rho_a=\sum_b m_b W^s_{ab}(h_a,n_a)
\label{density}
\end{equation}

\begin{equation}
\begin{split}
\frac{d\bf v_a}{dt}=-\sum_b m_b \left[\frac{P_a}{\Omega_a\rho_a^2}{\bf\nabla} W^s_{ab}(h_a,n_a)+\right.\\
\left. +\frac{P_b}{\Omega_b\rho_b^2}{\bf\nabla} W^s_{ab}(h_b,n_b)+ 
\Pi_{ab}{\bf\nabla}\widetilde W^s_{ab}\right]\,.
\end{split}
\label{momentum}
\end{equation}

\begin{equation}
\begin{split}
\frac{du_a}{dt}=\sum_{b=1}^{n_b}m_b({\bf v}_a-{\bf v}_b)\cdot\left(\frac{P_a}{\Omega_a\rho_a^2}{\bf \nabla}W^s_{ab}(h_a,n_a)\right.\\
\left.+\frac{\Pi_{ab}}{2}{\bf\nabla}\widetilde W^s_{ab}\right.)\,,
\label{energy}
\end{split}
\end{equation}

\noindent
where $W_{ab}^s$ is given by Eq.(\ref{sinc}), $\widetilde W^s=0.5(W^s(h_a,n_a)+W^s(h_b,n_b))$ and the remaining symbols have their standard meaning \citep{monaghan05}. The parameter $\Omega_a$ includes the relevant information to compute not only the grad-h, but also the grad-n derivatives:

\begin{equation}
\begin{split}
\Omega_a=1-\left[\left(\sum_b m_b\frac{\partial W_{ab}(h_a,n_a)}{\partial h_a}\right)~\left(\frac{\partial h}{\partial\rho}\right)_a+\right.\\ 
\left.\left(\sum_b m_b\frac{\partial W_{ab}(h_a,n_a)}{\partial n_a}\right)~\left(\frac{\partial n}{\partial\rho}\right)_a\right]\,,
\label{omega}
\end{split}
\end{equation}

\noindent
where $(\frac{\partial h}{\partial\rho})_a=-\frac{h_a}{d~\rho_a}$. The last term on the RHS in Eq.(\ref{omega}) accounts for the 
correction for the gradient of the exponent of the $sinc$ kernel $n$, which  the distinctive feature of our proposal. The derivative 

\begin{equation}
\left(\frac{\partial n}{\partial\rho}\right)_a=\frac{\partial n_a}{\partial\xi_a}\cdot\frac{\partial\xi_a}{\partial\lambda_a}\cdot\frac{\partial\lambda_a}{\partial\rho_a}
\label{dndro}
\end{equation}

\noindent
can be computed from Eqs. (6), (8), and (9). Details of the implementation of the grad-h and grad-n corrections  are given in Appendix~A.

For the artificial viscosity (AV) we used the recipe described in \citet{monaghan97}, inspired by the Riemann solvers formulation, where the term $\Pi_{ab}$ accounting for the viscous pressure is
\begin{equation}
\Pi_{ab}=\begin{cases}
-\frac{\alpha}{2}\frac{v^{sig}_{ab}~w_{ab}}{\bar\rho_{ab}} & \text{for ${\bf r}_{ab}\cdot {\bf v}_{ab} < 0$}\,,\\
0& \text{otherwise}\,,
\end{cases}
\label{avis}
\end{equation}

\noindent
where $v^{sig}_{ab} = c_a +c_b-3~w_{ab}$ is an estimate of the signal velocity between particles $a$ and $b$~ and is given by $w_{ab}={\bf r}_{ab}\cdot {\bf v}_{ab}/\vert {\bf r}_{ab}\vert$ is the relative velocity projected onto the separation vector. Following \citet{springel2010b}, we used a constant $\alpha=4/3$ to carry out the simulations described below, so that $\Pi_{ab}$ remains close to the classical SPH artificial viscosity introduced by \citet{monaghan1983}. This particular form of the AV has the advantage that there is no explicit dependence of viscosity on the smoothing length, because using $n({\bf r}, t)$ makes $h$  less reliable as an indicator of resolution. In principle, the viscous terms in Eqs. (\ref{momentum}, \ref{energy}) could be computed using a different kernel index than those depending on gas pressure. We have not found relevant differences among the results of the tests described below when the actual index $n_a$ of the particle, calculated with expression (\ref{indexes}), or the constant baseline value $n_0$ is used to compute the viscous part of Euler equations. The only exception was the 1D blast wave test, where the variable exponents $n_a(x,t)$ led to a narrower spike in the density peak. For that reason variable kernel indexes were also used to estimate the contribution of viscous terms to momentum and energy.

The calculation of the Euler equations is preceded by a brief {\it preconditioning stage}, where the optimal values of $h$ and $n$ are set. The value of $h$ is chosen  so that the mass within a volume $h^d$ is constant during the calculation. An initial pilot value of the density $\rho_a$, calculated with a trial $n_a$, as well as $\ln~\bar\rho_a$ are evaluated at this point. Then the self-consistent new values of $h_a$ and $n_a$ are found using the Newton-Raphson (NR) iterative scheme described in Appendix A. Regardless of setting $\Delta n=0$ in Eq. (\ref{indexes}) or imposing $\bar\rho=\rho$, the preconditioning algorithm is restored to the standard description, in which the kernel index is kept constant, and the smoothing length and density are jointly updated.  

The value of the free-parameter $\lambda_c$ sets the sensitivity of the kernel index with respect to $\lambda$. For $[n_0, \lambda_c, \Delta n]$ in Eq. \ref{indexes} we used $[5, 0.5, 5]$ in all the tests below, which yielded satisfactory results. A lower value of $\lambda_c$ leads to larger exponents, but also increases the noise level.

\section{Hydrodynamic tests}
\subsection{One-dimensional tests}

\subsubsection{Blast waves}

Reproducing a strong 1D blast wave with a known analytical solution is a powerful test for any hydrocode. The main goal here is to analyze if the adaptive kernel index algorithm is robust and leads to results better than or at least comparable with the calculation with constant exponents. Our first test was carried out with the same initial setting as in \cite{monaghan97}. From now on, initial models are specified by $[\rho, vel, \gamma, P, \Delta]$, where $\gamma$ is the constant that relates pressure and internal energy, $P=(\gamma-1)\rho u$, and $\Delta$ the interparticle distance. For this test $[1, 0, 1.4, 10^3, 0.005]$ for $x<0$ and $[1, 0, 1.4, 0.01, 0.005]$ for $x>0$. Simulations were carried out using constant kernel indexes $n=3$ and $n=5$, as well as variable kernel indexes. A model was also run with variable exponents $n_a$, but keeping $n_a=n_0$ in the viscous terms of momentum and energy equations. 

The results of the calculations are summarized in Fig.~\ref{figure5}, where we show the profiles of density, velocity, and kernel index at $t=0.08$~s. There are no substantial differences between the different models. They all agree well with the analytical profile. From the fine details, however, we see that the calculations with  $n(x,t)$ depict the density in the rarefaction tail of
the wave slightly better (bottom left panel in Fig.~\ref{figure5}). In this case, we see a small spike that only affects one particle in the plateau at highest density. This feature disappears if constant $n_0$ is taken to compute $\nabla W_{ab}$ in the viscous terms of Eqs.~(\ref{momentum}), (\ref{energy}). 

The  algorithm to self-adapt $n(x,t)$ is robust and works very well, detecting strong gradients of density and interphases, as suggested in the bottom right panel of Fig.~\ref{figure5}. The index of the $sinc$ kernels changes only in a very narrow region at the sides of the density peak, almost reaching its highest allowed value $n=10$. Note also the similarities with the mountain-like~static profile of Fig.~\ref{figure4}, where $n(x,t)$ peaks twice around the maximum in the  density profile. The density profiles of models with different $n$ are also similar at the low-density tail in the shocked region.    

As a  variation of the previous test, we tracked the evolution of a 1D point-like explosion. In this case, the density contrast between the peak and the bottom of the profile is higher than in the preceding case. We started from a homogeneous distribution of particles with $[1, 0, 1.4, 0.01, 5~10^{-4}]$.  The explosion was initiated by increasing  the internal energy of the central particle by a factor $10^6$. As before, the evolution was followed using three prescriptions for the kernel index, $n=3, n=5$ , and $n$ adaptive. Figure~\ref{figure6} depicts the density profile at time $t=0.016$~s for the different indexes and initial resolution $h_0=1.5\Delta$. The pattern consists of two strong shock-waves moving in opposite directions, separated by a diluted region. Again we see that $n$ changes abruptly around the discontinuities. Nevertheless, the density profile  matches well, regardless of the value chosen for the kernel index. The exception is the central diluted zone, which is better described when equalization is turned on, as shown in the bottom left panel of Fig.~\ref{figure6}. The profile of pressure (normalized to the pressure peak at the shock front) is also depicted in the same figure and compared with the analytical profile. Around the peak of the blast all cases agree well, but this is different at the central, low-pressure region. Still, the calculation with adaptive index provides a better approach to the pressure in that zone. This feature is also seen in two dimensions, as commented in Sect.~ 5.2.1.

\subsubsection{Shock-tube test}

This is a similar test as before, but now the shock and the rarefaction waves are much weaker. In this case, a box is filled with a gas so that the pressure in the leftmost part of the box is higher than in the right side. At $t=0$~s both regions are separated by a wall. When the wall is removed, the two regions begin to mix and a shock wave appears that moves through the low-pressure region, while a rarefaction wave digs into the  high-pressure zone. The initial conditions are  left $[1,0,1.4,1,2.5~10^{-4}]$, right $[0.125, 0, 1.4,0.1, 2~10^{-3}]$.  

A summary of the results is given in Fig.~\ref{figure7} where the profiles of density, internal energy, pressure and velocity are shown and compared with the analytical values. In this test the resulting profiles calculated with constant $n=3$, $n=5$ and $n$ adaptive are nearly identical  and only the result for $n(x,t)$ is given. This matches the exact profile very well. Nevertheless, we also see a sharp spike in internal energy and pressure at the contact discontinuity. This feature (also present in the calculations with n=3, n=5) is known to show up when the contact discontinuity is not smoothed at t=0 s and there is no heat diffusion term, driven by an artificial conductivity, included in the energy equation.

In Fig.~\ref{figure8}, we show the evolution of the profile of the kernel index $n(x,t)$. The highest values of $n$~ are achieved at $t\simeq 0$ when the  density contrast around the contact discontinuity is highest and its profile steep. Nevertheless, they decay fast to values close to the baseline value $n_0=5$ as soon the self-similar state is achieved, which makes the results very similar to those obtained with constant $n$.

\subsubsection{Sj\"ogreen test}

As described by \cite{einfeldt91}, this gas-dynamics problem involves the propagation of two symmetric rarefaction waves through a perfect gas with $\gamma=1.4$. The Sj\"ogreen test can be easily handled with SPH, but not with methods using iterative Riemann solvers unless special techniques are used. To initiate these waves, the initial conditions were set as in \cite{monaghan97}, with one half of the system moving to the right with $[1, 2, 1.4, 0.4, 0.001]$, while the other half moves to the left with $[1, -2, 1.4, 0.4, 0.001]$. As a result, a cavity filled with a very diluted gas appears at the center of the system. The geometry of this fluid cavity resembles the valley-like profile depicted in the upper left panel of Fig. \ref{figure4}. A comparison between the profiles of several magnitudes at $t=0.9$~s, obtained with and without equalization, is provided in Figs.~\ref{figure9} and \ref{figure10} for two values of the initial smoothing length $h_0$ (note that a logarithmic scale was used to highlight the differences in the diluted region).  Even though the results are good in all cases, there is a clear improvement when the  equalization algorithm is included, especially for $h_0=3\Delta$. The lowest values achieved by the density, internal energy, and pressure are closer to the analytical expectations. These results agree qualitatively with the static valley-like case of Fig.~\ref{figure4}. The velocity profile at the center is slightly flatter when the adaptive kernel index is used, being also closer to the analytical solution. 

The profiles of $h$ and $n$ for the case $h_0=1.5\Delta$ are shown in Fig.~\ref{figure10}. Both the smoothing length and the adaptive kernel index steeply increase in the vicinity of the density minimum. The variable index $n$ is thus controlling the loss of resolution caused by the growth of $h$. This effect is not linear, however, because the model calculated with equalization has a higher value of $h$ at the lowest density than models calculated with constant $n$. This is a consequence of the strong coupling between $\rho, h,$ and $n$ which, as mentioned above,  are self-consistently found  at each step using an iterative  Newton-Raphson (NR) scheme.

\subsection{Multidimensional tests}

\subsubsection{Sedov test in 2D}

To study the evolution of a spherical Sedov-Taylor blast wave,
we conducted a test that involve the propagation of a delta-function signal. This gas-dynamical problem includes a point-like explosion inside a homogeneous system. The explosion rapidly evolves towards a self-similar wave with a known analytical solution \citep{sedov1959}. This is a very demanding test in more dimensions than 1D, where the resolution is usually too low to yield reliable values of the magnitudes around the peak of the wave or close to the origin of the explosion. We wish to know if the combined adaptive $h - n$ scheme can describe this phenomenon better. To trigger the explosion, a $\delta$-like  function  was imposed  on the internal energy at t=0 s,

\begin{equation}
u(r)=u_0\exp\left[\frac{-r^2}{\sigma^2}\right]
,\end{equation}

\noindent
where $r$ is the distance to the explosion center and $u_0=10^7$~erg.g$^{-1}$, $\sigma=0.02$~cm. For this test the initial interparticle separation was $\Delta= 4~10^{-3}$. The initial value of the smoothing length was set to encompass ${\mathcal N}= 46$ neighbors. The profiles of several magnitudes during the self-similar evolution of the blast are shown in Fig.~\ref{figure12}. As in the preceding tests, the equalization mostly affects the shocked region, although its imprint is not strong. The rarefacted tail of the blast wave is better described when variable $n({\bf r},t)$ are used. In particular, the pressure profile downstream shows a clear dependence on the index of the $sinc$ kernels. The conservation of energy is quantified in the bottom right panel of Fig.~\ref{figure12} with the adaptive $n({\bf r},t)$ scheme providing the best results. It is interesting to note that although the SPH formalism is built to exactly conserve energy, the conservation is usually  not perfect in practice owing to the small errors in particle localization, especially when neighboring particles have very different smoothing lengths. Using a large $n$~in $S_n$~makes interpolations less susceptible against small fluctuations at the outer edge of the kernel. In this sense, including the equalization lowers these errors and improves the total energy conservation. The profile of $n(\bf{r},t)$, when equalization was included, is depicted in the bottom left panel of Fig.~\ref{figure12} and in Fig.~\ref{figure13}. We see two regions where the kernel index becomes higher than its baseline value $n_0=5$, one around the shock front peaking at $n\simeq 6.5$ and other at the post-shock diluted region with a highest value of $n\simeq 9.5$.

Finally, a calculation was launched with a high value of the kernel exponent, $n=10$, in all particles. A color density map  for cases $n=3$, $n=5$, $n=10$ and $n$ adaptive is provided in Fig. \ref{figure14}. For $n=10$ the density distribution is affected by the initial particle setting in a rectangular lattice (sometimes referred to as hour-glass instability). As expected, the spherical symmetry is better preserved for the low-order interpolator $n=3$, but the cases $n=5$ and $n$ adaptive~are also very good. We conclude that the use of high kernel indexes must be reserved to fluid regions that host sharp density gradients (see also Sect. 5.2.2).  Low-order interpolators are more efficient in  suppressing numerical noise, but they are less accurate and more prone to undergo pairing instability. A conservative procedure is to take a moderate index $n$ in all fluid regions to reduce the numerical noise, but switch to a larger $n$ wherever a discontinuity is found.

\subsubsection{Kelvin-Helmholtz instability in 2D}

The Kelvin-Helmholtz instability appears when there is a sufficient shear velocity in the interface layer between two fluids with different densities. Small perturbations of the velocity field in the orthogonal direction to the interface  emerge and lead to a mixing of the two fluids. This is usually simulated in a box with periodic boundary conditions, where two fluid regions are defined with densities $\rho_1$ and $\rho_2$ . The two layers have opposite parallel velocities, which leads to a shear discontinuity in the contact interface. To develop the instability, a small perturbation is seeded at the interface as a sinusoidal mode of length scale $L$. Recent SPH simulations of the KH instability can be found in \citet{mcnally12} and \citet{hopkins13}.

We simulated a central band of a high-density fluid $\rho_1$ moving in a low-density medium $\rho_2$ in a squared lattice of $1$ cm side in the XY plane using $N=62,500$. The density around the interface was not smoothed. The initial setting was $[1, -0.5, 5/3, 2.5, 0.006]$ for $y\leq 0.25, ~y\geq 0.75$ and $[2, +0.5, 5/3, 2.5, 0.003]$ for $0.25< y < 0.75$. The initial smoothing-length was chosen so that every particle sees $\mathcal N=50$ neighbors.      

A sinusoidal perturbation of the $v_y$ component of the velocity field was seeded at $t=0$. Then, for the initial velocity we have
\begin{equation}
 v_y(x)=\Delta v_y\sin{(m\pi x)}\,,
\end{equation}

where we took $m=2$ and $\Delta v_y=0.01$ cm.s$^{-1}$, a small perturbation indeed. 

First of all, we would like to stress that the calculation with $n=3$ was a complete failure because the perturbation failed
to emerge.  The reason for this was that according to Fig.~\ref{figure3}, the initial number of neighbors is much higher than necessary to suppress pairing instability. In the calculations with  $n=5$, however, the pair $n=5,~\mathcal N=50$, lies only moderately above that line, and no trace of particle clustering was detected
during the simulation. Particle clustering can also be avoided, even for $n\simeq 3$,~using a a different SPH approach to the fluid equations, such as those based on an integral approach to the derivatives (IAD) \citep{garciasenz12}.

Figure~\ref{figure15} shows a density color map of the growth of the Kelvin-Helmholtz instability at different times for the calculations using $n=5$, $n=10$ and $n$ adaptive. The simulation with $n=10$ is manifestly poorer, suggesting again (see the preceding section) that choosing a large $n$ from the beginning is not a safe option. Cases $n=5$ and $n$ are almost indistinguishable; both lead to a clear growth of the instability with its characteristic pattern. The last row of Fig.~\ref{figure15} shows a color map of $n ({\bf r},t)$ for the same models as depicted in row 3. The tracking algorithm perform well because $n$ only increases in a thin shell around the interface. Nevertheless, the change in $n$ is not large and the dynamical evolution remained close to that with $n=5$.

\subsubsection{Astrophysical application: Gravitational collapse of a polytrope}

Finally, we simulated the gravitational collapse of a Sun-like polytrope  with and without the equalization algorithm and compared the results  with the output of a well-known 1D Lagrangian hydrocode \citep[the AGILE hydrocode by][]{liebendorfer02}. We carried out the 1D models with AGILE taking  260 grid points.  This   
provides an output with much better resolution than that of the SPH hydrocode and serve as a suitable reference model.  

The initial model used for the comparison was a $1M\sun$ spherically symmetric polytrope of index 3. The radius was set to $1R\sun$, which results in a central density of $\rho_c= 76$ g.cm$^{-3}$. We built specific equilibrium initial models for each case (with and without equalization) by distributing $N=10^5$ particles in 3D according to the 1D density profile, and let them relax to the hydrostatic equilibrium. The EOS used in the simulations was that of a perfect gas with $\gamma=5/3$. The initial value of $h$ in the SPH calculations was chosen to encompass $\mathcal N=100$ neighbors. 

The equilibrium structure was then suddenly destabilized by removing  $20\%$ of its internal energy, so that the star collapsed under the force of gravity. At some point, the collapse in the central zone was halted because of the increase of pressure,  and an accretion shock formed that moved through the infalling material to ultimately eject the surface layers of the polytrope. That scenario contains several pieces of physics of great interest because accretion shocks and pulsational instabilities are very common in astrophysics. 

The evolution of the central density during the collapse and the rebound of the star is shown in Fig.~\ref{figure16}, the profile of several variables at three elapsed times is provided in Fig.~\ref{figure17}. In general, all calculated models show a similar behavior during the implosion and first oscillation of the polytrope. The first peak of central density is achieved after $\simeq 15$ minutes in all simulations. Nevertheless, the exact value of the peak is affected because the resolution is
higher in the AGILE 1D calculation and lower for SPH with $n=3$, as expected. The calculations with $n=5$ and $n$ adaptive virtually led to the same maximum in the central density. The discrepancy between $n$ adaptive and the reference 1D model is $\simeq 6\%$. Fig.~\ref{figure16} also depicts the evolution of the fraction of total energy lost during the first hour, which remains below $0.2\%$ for all the SPH models. As in the Sedov test, the calculations with higher exponents, $n=5,~n({\bf r}, t)$ conserve the energy better than that with the cubic-spline-like kernel, case n=3.

The profiles of density, velocity, and internal energy at times $t=870$~s, 1086~s, and 1311~s are depicted in Fig.\ref{figure17}. The density profiles do not show any significant difference between the $n=5$ and $n({\bf r}, t)$ calculations. In  both cases the discontinuity at the accretion shock is smoothed in a similar way and is less pronounced  than in the reference model. The radial velocity profiles are shown in the upper right panel of Fig.\ref{figure17}. They show some differences at the position of the accretion shock; the simulation using $n({\bf r}, t)$ better matches the AGILE results at $t=1086$~s and $t=1311$~s. In particular, the lowest velocity is much better captured when the equalization is included. A similar behavior is observed in the profile of the specific internal energy depicted in the bottom left panel of Fig.\ref{figure17},~but there the differences are not as accentuated as in the velocity profile. The bottom
right panel of the same figure shows the distribution of $n$ along the star. The algorithm  detects both the accretion shock and the surface of the polytrope at $t=870$~s, while for longer elapsed times $n({\bf r}, t)$ follows a wall-like profile with the baseline value $n_0=5$ until $\simeq 0.4$ $R_{\sun}$ and $n\simeq 10$ at the surface.

\section{Discussion and conclusions}

In the standard formulation of the SPH method the resolution is bounded to the local density  value, meaning that the rarefacted zones of the fluid are intrinsically handled with a lower resolution than the high-density regions. We proposed a method to equalize the error in diluted regions that is robust and easy to implement, with a low computational overload. The formulation of the method relies on the definition of a local estimator of the linearity of density. According to Eq.~(\ref{estimator}), the definition of that estimator, $\lambda$, is fairly simple and its value is used to control the accuracy of the interpolations. A similar method was proposed by \cite{sigalotti06} as a way to set the value of the smoothing length at each step-time. Our proposal differs from that of these authors in several ways. First, in our method $h$ is set in the standard manner, keeping the mass constant around a particle, while the value of $\lambda$ sets the value of the exponent, $n$, of the $sinc$ kernels. Second, unlike \cite{sigalotti06},  who neglected the grad-h terms, we included the grad-h and grad-n corrections to the momentum and energy equations. The computation of these corrections is compatible with the Lagrangian derivation of the fluid equations. Third, the specific mathematical expressions used to set $n$ are different from those used by  Sigalotti and coworkers to set the value of $h$. In our proposal, we constrained  $n$ to the range $n_0\le n\le n_0+\Delta n$, with the boundaries achieved asymptotically. 

The proposed algorithm works well with static 1D particle distributions. According to Fig. \ref{figure4}, the zones with sharp density gradients are better described using an adaptive kernel index $n(x)$. Unlike the adaptive $h(x)$, the improvement due to $n(x)$ is  more pronounced in the low-density tail of the profiles. Therefore using both $h(x)$ and $n(x)$ tends to equalize the error along the system.

In hydrodynamic calculations, however, the improvements are not as pronounced  as in the static profiles. The main reason is that sharp density gradients are smoothed by the artificial viscosity that widens the discontinuities to twice or thrice the smoothing length. Moreover, the mechanism by which $h$ and $n$ self-adapt is not longer linear, and sometimes a self-consistent increase of $n$ is followed by an increase in the number of neighbors, making the enhancement in resolution less noticeable (see, for instance, Fig.~\ref{figure11}). Still, the hydrodynamic tests confirm the main results attained with 1D static profiles: a moderate improvement in the description of the rarefacted regions of the gas, usually attached at the rear tail of shock waves. The careful handling of these post-shock regions must not be disregarded because it is as important as the shock front itself: in these tails hydrodynamic instabilities may grow under the appropriate physical conditions (for example, the Rayleigh-Taylor instability in the regions between the forward and reverse shocks in supernova remnants).        

The search for the optimal $n({\bf r},t)$ can be made in the same NR loop as was used to update $h({\bf r},t)$ with very little changes. At each iteration the value of $\sum_b m_b \frac{\partial W_{ab}}{\partial n_a}$ and the local arithmetic mean of $\ln\rho$ have to be stored, but the overload is small if a list of the neighbors of each particle is stored in an array and used to localize particles when necessary. A switch can be used to include
or exclude the equalizing option, as shown in Fig.~\ref{AppendixFig1}. The algorithm is very efficient in detecting discontinuities. It was able to track the contours of shocks, walls, and surfaces in all the tests. The equalization does not interfere with the development of the Kelvin-Helmholtz instability either because it neither enhances nor diminishes the growth.      

The application to a specific astrophysical problem: the collapse and subsequent rebound of a Sun-like polytrope was also satisfactory. The calculation with equalization led to better profiles of velocity and specific internal energy with an adaptive $n({\bf r},t)$ increasing in the rear of the accretion-shock front and at the surface. Nevertheless, while a high value of $n$ at the shock is driving  a clear enhancement of the internal energy and radial velocity profiles, its impact on the surface layers was weak. The reason is that interpolations at the boundaries of self-gravitating bodies are not as accurate as in the interior because of the scarcity of sampling points in the outermost regions of the envelope. To adequately solve the surface layers in 3D and estimate the real effect of equalization a huge increase in the number of particles would be necessary. 

Among other advantages, the $sinc$ family of interpolators introduces an additional degree of freedom to control the resolution in SPH. The simultaneous (implicit or semi-implicit) search for $h$ and $n$ increases the computational burden, but this is no great concern unless
very many particles require a hard refining of $n$ . In this respect, we estimated a $\simeq 10 \%$ overload in the simulation of the 2D Sedov point-like explosion. The computational penalty will be weaker in current astrophysical scenarios where gravity and/or a complex physics are incorporated in the numerical scheme. 

A priori, working with $n({\bf r}, t)$ can also be a potential source of numerical noise, which may affect the development of small fluid instabilities. In this respect, we found no spurious effect in the growth of the KH instability, but more work is needed to confirm this last point. Additionally, other functional forms of the estimator $\lambda_a$, different from that used in this work given by Eq.~(\ref{estimator}), might be devised to control $n$ and better adapt the abilities of current SPH codes to handle specific physical problems.

\section*{Acknowledgements}

This work has been funded by the Spanish MEC grants AYA2010-15685, AYA2011-23102 and DURSI of the Generalitat of Catalunya (D.G.S. and J.A.E.). RMC acknowledges the support by the Swiss Platform for High-Performance and High-Productivity Computing (HP2C) within the {\em supernova} project and the Platform for Advanced Scientific Computation (PASC) within the DIAPHANE project. RMC and KE were also supported by the ERC grant FISH. D.G.S. was also supported by the EuroGENESIS and CompStar progams. The rendered SPH plots were made using the freely available $SPLASH$ code \citep{price07}.


\Online
\begin{appendix} 
\section{Implementation of the algorithm to compute grad-h and grad-n}

A suitable mathematical expression giving the corrections by the grad-h and grad-n terms can be obtained as a simple extension of the reasoning used to compute the grad-h terms \citep{rosswog09}. The discretized fluid movement equations are derived using the Euler-Lagrange formulation

\begin{equation}
\frac{d}{dt}\left(\frac{\partial L}{\partial {\bf v_a}}\right)-\frac{\partial L}{\partial {\bf r_a}} = 0, 
\label{Appeuler_lagrange}
\end{equation}

\noindent where ${\bf r_a}$ and ${\bf v_a}$ refer to the position and velocity of particle $a$. The Lagrange function of the system is

\begin{equation}
L = \sum_b \frac{1}{2} m_b \left [v_b^2+ u_b (\rho_b, s_b)\right]\,
\label{Applagrangian}
,\end{equation}

\noindent where $u_b, s_b$ are the specific internal energy and entropy of particle $b$. Inserting Eq.~(\ref{Applagrangian}) into Eq.~(\ref{Appeuler_lagrange})~and admitting isentropic evolution, $\partial u_b/{\partial\bf r_a}= P_b~\rho_b^{-2}~\frac{\partial \rho_b}{{\partial\bf r_a}}$ the movement equations for particle $a$ are written 

\begin{equation}
m_a \frac{d{\bf{v_a}}}{dt}= - \sum_b m_b~\frac{P_b}{\rho_b^2}\frac{\partial{\rho_b}}{\partial{\bf r_a}}.
\label{Appeqmovement}
\end{equation}

Following \cite{rosswog09}, the density gradient in Eq.~(\ref{Appeqmovement})  (also needed to compute the energy equation) is calculated as 

\begin{equation}
\begin{split}
\frac{\partial\rho_b}{\partial{\bf r_a}}=\sum_c m_c\left[\nabla_a W_{bc}(h_b, n_b)+\frac{\partial W_{bc}(h_b,n_b)}{\partial h_b}\frac{\partial h_b}{\partial\rho_b}\frac{\partial\rho_b}{\partial{\bf r_a}}+\right.\\
\left. \frac{\partial W_{bc}(h_b,n_b)}{\partial n_b}\frac{\partial n_b}{\partial\rho_b}\frac{\partial\rho_b}{\partial{\bf r_a}}\right]=\frac{1}{\Omega_b}\sum_cm_c \nabla_a  W_{bc}(h_b, n_b)\,,
\end{split}
\label{Appgradrho}
\end{equation}

\noindent where $\nabla_a$ is the derivative with respect to
the spatial coordinates and   

\begin{equation}
\begin{split}
\Omega_b=1-\left[\left(\sum_c m_c\frac{\partial W_{bc}(h_b,n_b)}{\partial h_b}\right)~\frac{\partial h_b}{\partial\rho_b}+\right.\\ 
\left.\left(\sum_c m_c\frac{\partial W_{bc}(h_b,n_b)}{\partial n_b}\right)~\frac{\partial n_b}{\partial\rho_b}\right]\,,
\label{Appomega}
\end{split}
\end{equation}

\noindent with $\frac{\partial h_b}{\partial\rho_b}=-\frac{h_b}{d\rho_b}$ and  

\begin{equation}
\frac{\partial n_b}{\partial\rho_b}=\frac{\partial n_b}{\partial\xi_b}\cdot\frac{\partial\xi_b}{\partial\lambda_b}\cdot\frac{\partial\lambda_b}{\partial\rho_b}\,,
\label{Appdndro}
\end{equation}

\noindent which can be computed from Eqs.~(\ref{indexes}), (\ref{ffunction}) and (\ref{Derivestimator}). Inserting Eqs. (\ref{Appomega}) and (\ref{Appgradrho}) into Eq.~(\ref{Appeqmovement}),~the form of the momentum equation used in this work, Eq.~(\ref{momentum}),~is easily recovered.  

The ensuing algorithm to update $h, n$ and calculate the grad-h, grad-n corrections of particle $b$ was implemented using a Newton-Raphson iterative scheme,

\begin{equation}
\left\{\begin{array}{rclcc}
G_b^1=\frac{C_b}{h^d}-\sum_c m_c W_{bc}(h_b, n_b)\\
G_b^2= n_b-n_0-\Delta n\cdot f(\xi_b)\,,
\end{array}
\right.
\label{G1G2}
\end{equation}

where $C_b=\rho_{b,0}~h_{b,0}^d$ is a constant, set at the beginning of the simulation, and $f(\xi_b)$ is defined in Eq. (\ref{ffunction}), hence,  

\begin{equation}
\left\{\begin{array}{rclcc}
\delta G_b^1=-\left(\frac{d~C_b}{h_b^{d+1}}+\sum_c m_c\frac{\partial W_{bc}(h_b, n_b)}{\partial h_b}\right)\delta h_b-\qquad\qquad\qquad\qquad\qquad\qquad\\
\left(\sum_c m_c\frac{\partial W_{bc}(h_b, n_b)}{\partial n_b}\right)\delta n_b
\qquad\qquad\qquad\qquad\qquad\qquad\qquad\qquad\\
\delta G_b^2= -~\Delta n\frac{\partial f(\xi_b)}{\partial h_b}\delta h_b+\left(1-\Delta n\frac{\partial f(\xi_b)}{\partial n_b}\right)\delta n_b\,.\qquad\qquad\qquad\qquad\qquad
\end{array}
\right.
\label{dG1dG2}
\end{equation}

Note that for $\Delta n=0$ the NR reduces to the standard scheme, where the kernel exponent is kept constant and h is updated according to the local density. For [$n_0, \Delta n, \lambda_c]$ we have taken $[5, 5, 0.5]$, which led to reliable results in the numerical tests with $\mathcal N\simeq 50, 100$ neighbors in 2D and 3D, respectively. To speed up the calculations, it is highly recommended to store the values of $sinc(\frac{\pi}{2}v), \frac{\partial{sinc(\frac{\pi}{2} v)}}{\partial v}, \ln[sinc(\frac{\pi}{2}v)]$ (the $\ln[sinc]$ can be used to compute $\frac{\partial W}{\partial n}$) as a function $0\le v\le 2$ in an array, and interpolate from them to obtain any kernel-related magnitude. A sample of $2\cdot 10^4$ equally spaced points was good enough for all tests presented in this work.

A flow chart of the preconditioning moduli is given in Fig. \ref{AppendixFig1}. When implementing the algorithm, particles that have already converged need to be carefully removed from the general NR loop, taking them into account only to compute $\bar\rho$. If the algorithm is well balanced and optimized, the computational overload should remain at a few $\%$, unless very many particles require hard refining.

\end{appendix}

\clearpage
 
\clearpage
\includegraphics[angle=0, width=\textwidth]{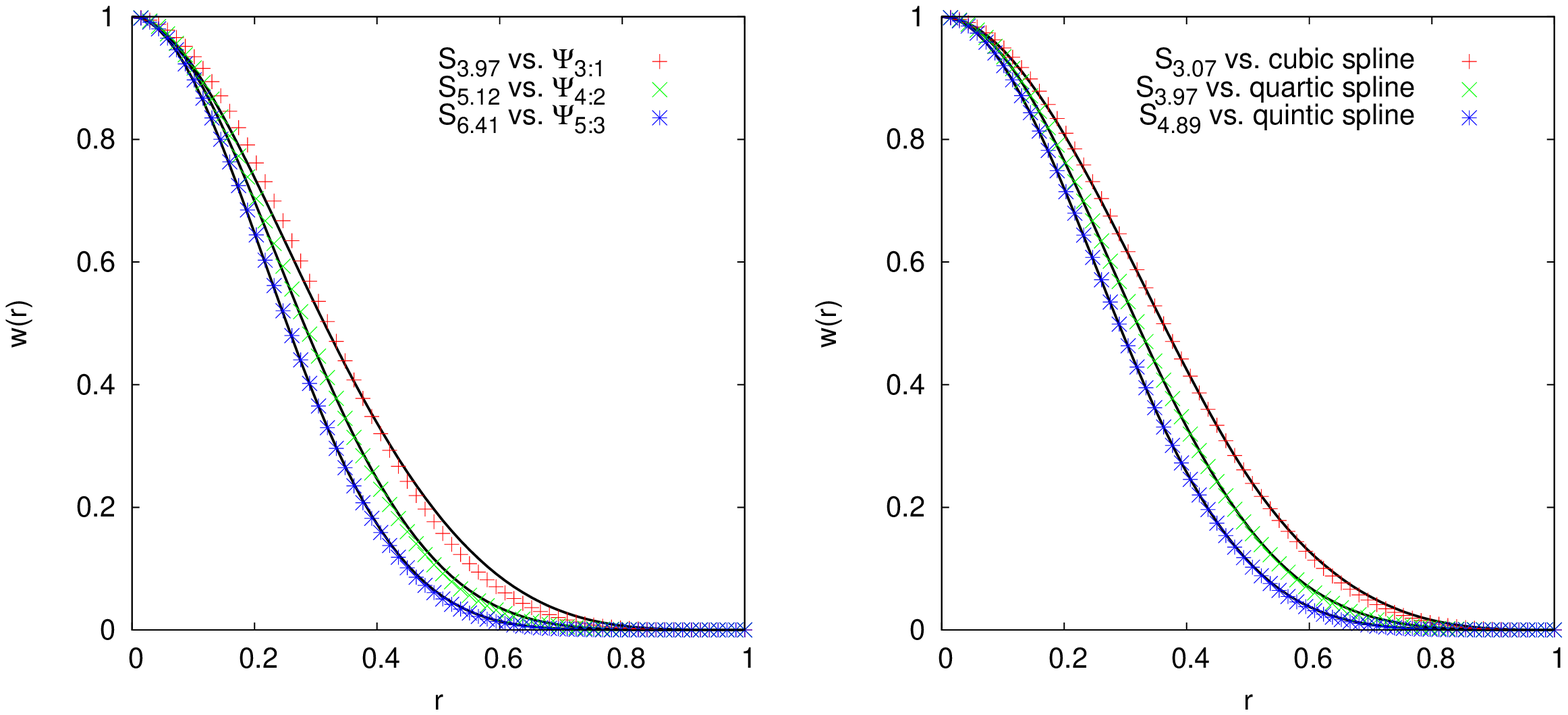}
\begin{figure}
\caption{Left: profile of several $sinc$ ($S_n$, continuum lines) and Wendland kernels ($\psi_{l:k}$) (rescaled to a common range [0,1]). Points  $+$ (in red), $\times$~(in green) and $\ast$~(in blue) are for $\psi_{3:1}$,   $\psi_{4:2}$~and  $\psi_{5:3}$~respectively. Right: same as before, but for $sinc$ and cubic, quartic, and quintic spline kernels.}

\label{figure1}
\end{figure}
\clearpage
\includegraphics[angle=0, width=\textwidth]{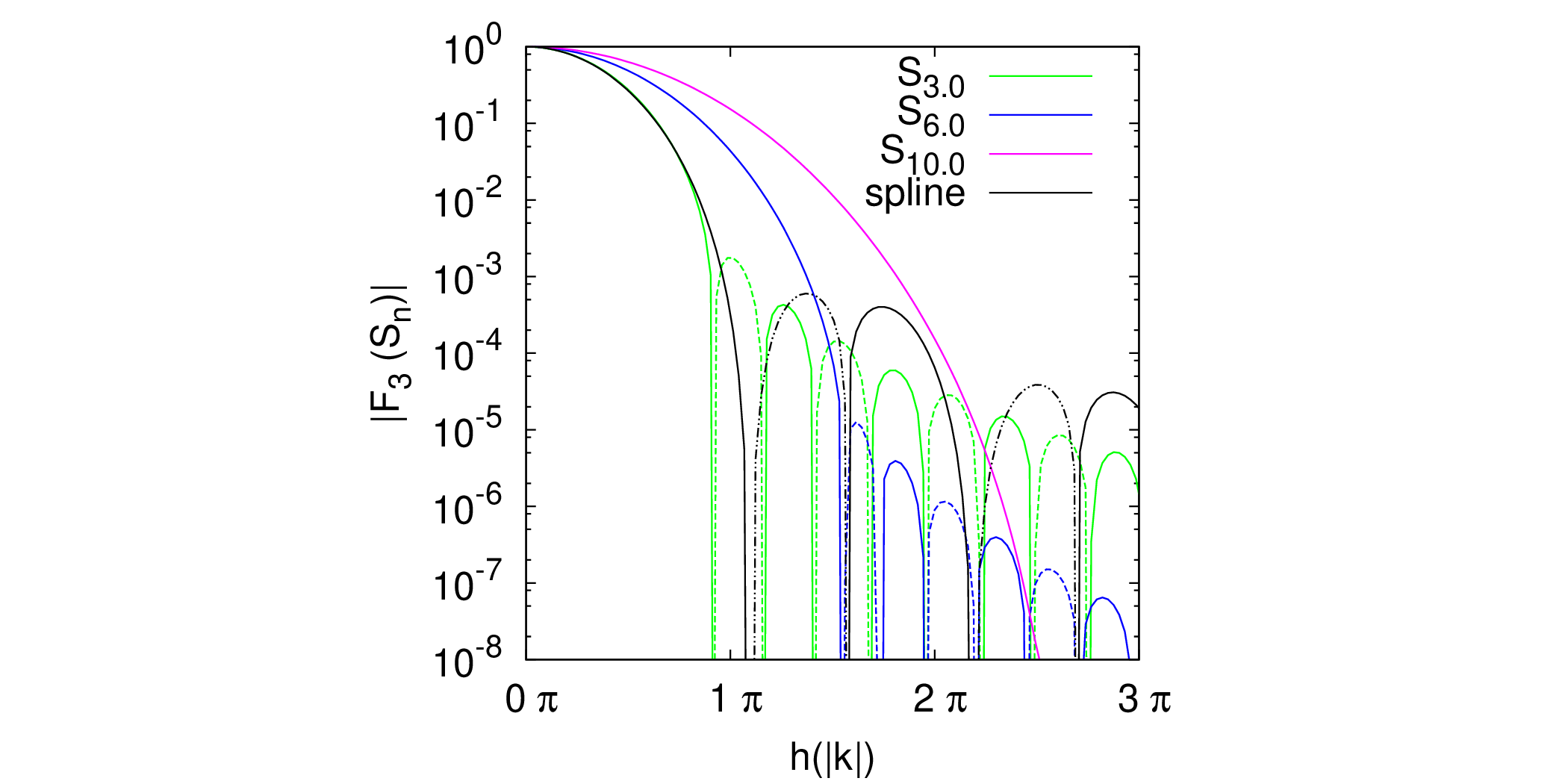}
\begin{figure}
\caption{Fourier transform, $\vert\mathcal F_3\vert$ of  $sinc$ kernels with $n = 3, 4$ and $10$ as well as the cubic spline, where the dashed lines indicate the negative portions of the curves. They can be compared with the Fourier transform of several Wendland and spline kernels given in Fig.~2 by \citet{dehnen12}.} 
\label{figure2}
\end{figure}

\clearpage
\includegraphics[angle=270, width=\textwidth]{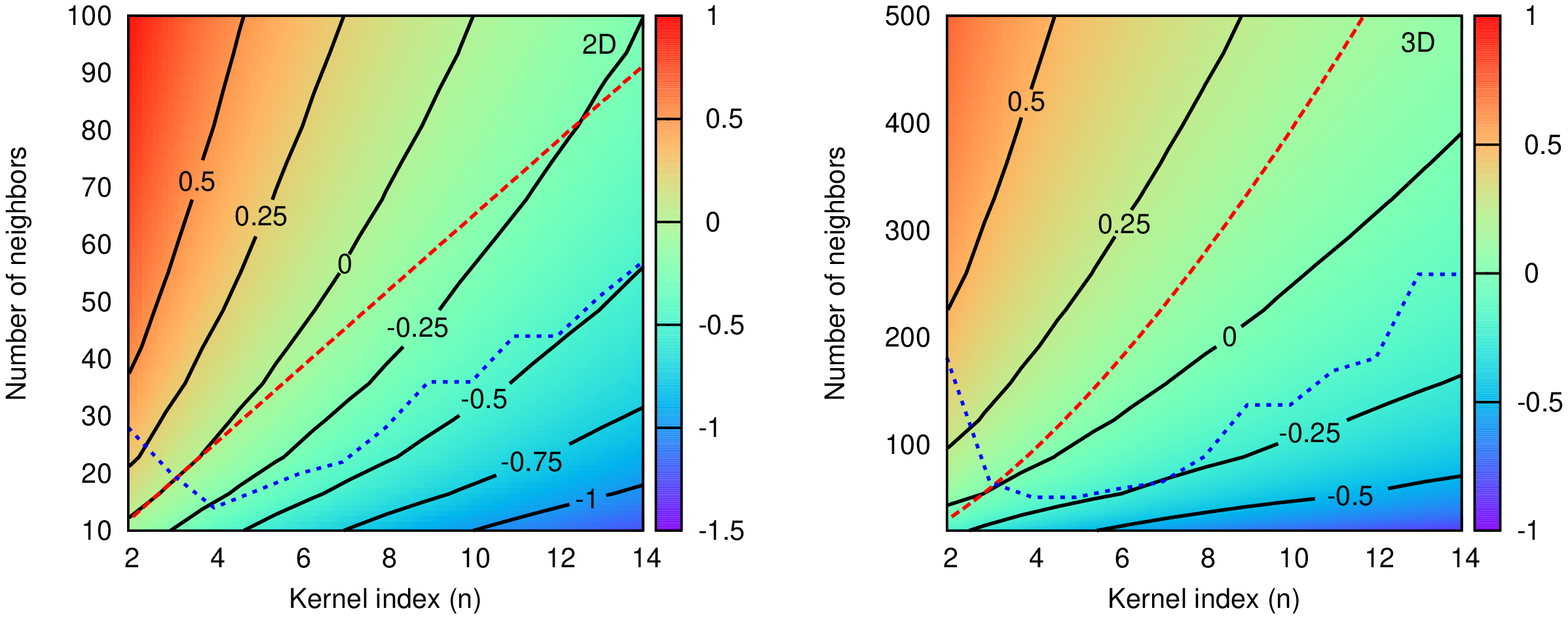}
\begin{figure}
\caption{Color map of the logarithm of the error in density estimation in 2D (left) and 3D (right) as a function of the kernel index and number of neighbors (see Sect.~3 for a complete explanation of this diagram). The dashed line in red is the rough critical limit separating the region susceptible to particle pairing (above the line). The blue dotted line denotes the region where the approach of integrals by summations in density calculation is too sensitive to particle distribution (below the line).}
\label{figure3}
\end{figure}

\clearpage
\includegraphics[angle=270, width=\textwidth]{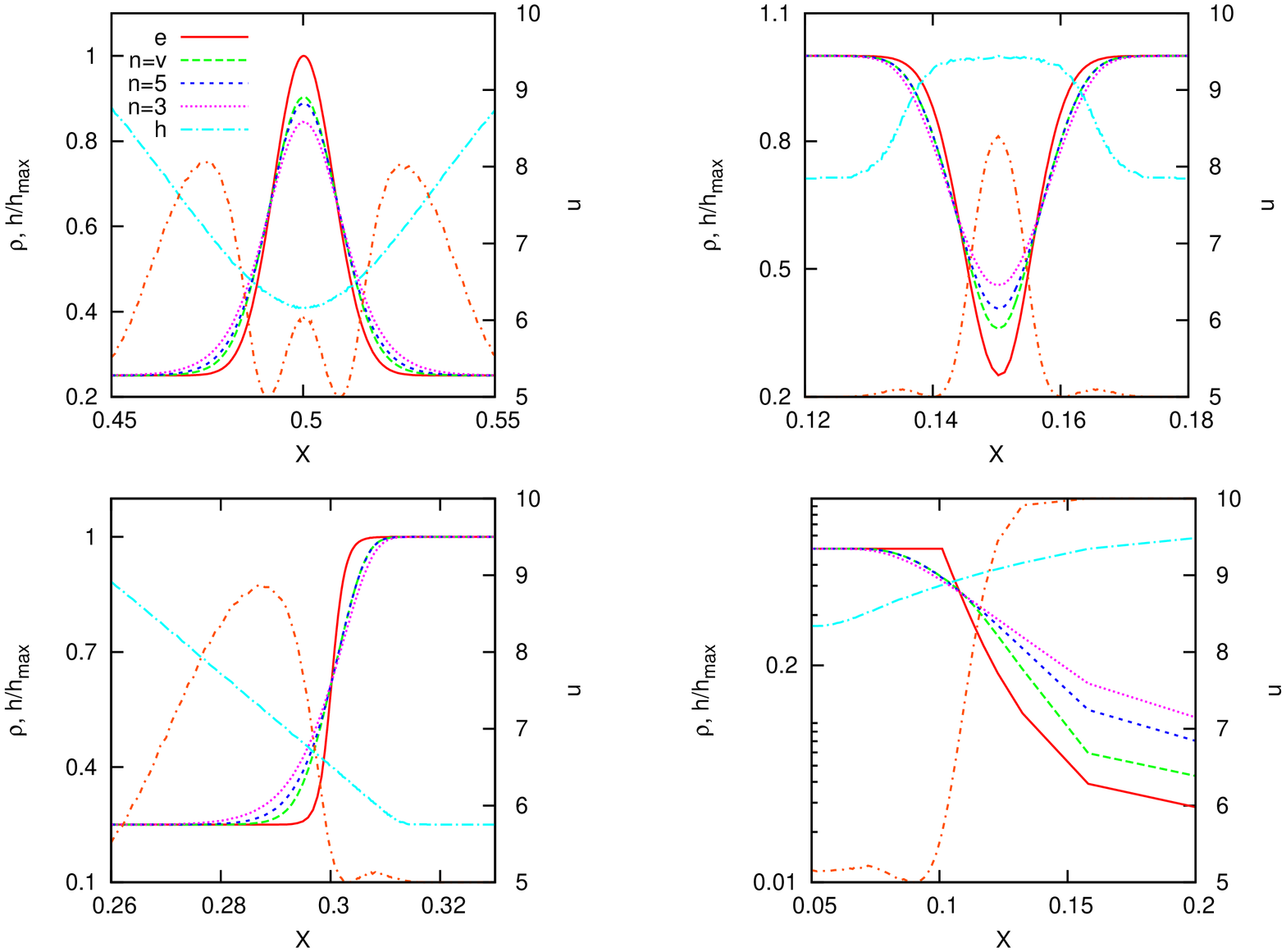}
\begin{figure}
\caption{Fitting of several 1D density profiles using the $sinc$ family of kernels with constant and self-adaptive indexes. The
upper
left panel shows a Gaussian, mountain-like, profile. The exact (e) analytical value is shown in red, the result with the variable (v) kernel index, constant $n=5$ and $n=3$ in green, blue, and pink, respectively. The light blue line shows the profile of the smoothing-length normalized to its highest value (achieved just at the limits of the system, x=0 and x=1). The orange line shows the profile of the kernel index $n$ associated with the green line. The same applies to the upper right (valley), the bottom left (wall), and the bottom right (cliff) profiles. } 
\label{figure4}
\end{figure}

\clearpage
\includegraphics[angle=-90, width=\textwidth]{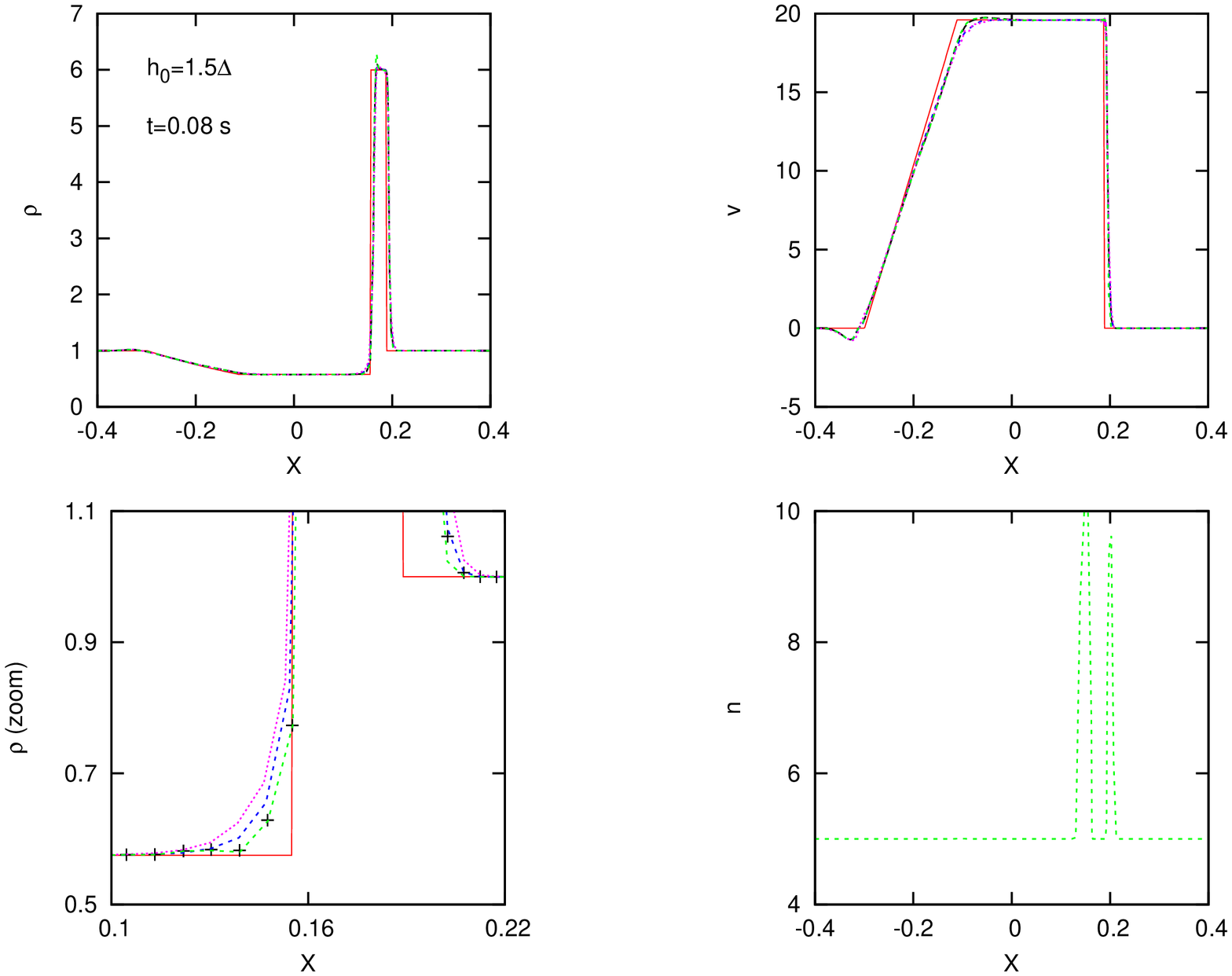}
\begin{figure}
\caption{Profiles of density, velocity, zoom of density and kernel index of the 1D blast-wave at time $t=0.08$~s. The continuum red line is the analytical value. Pink (dots), blue (dashed), and green (dashed) denote $n=3$, $n=5,$ and $n$ adaptive, respectively. The profile in black (dashed) lines and crosses (density zoom) plots $n$ adaptive, but keeping $n=5$ in the artificial viscosity terms.}
\label{figure5}
\end{figure}

\clearpage
\includegraphics[angle=-90, width=\textwidth]{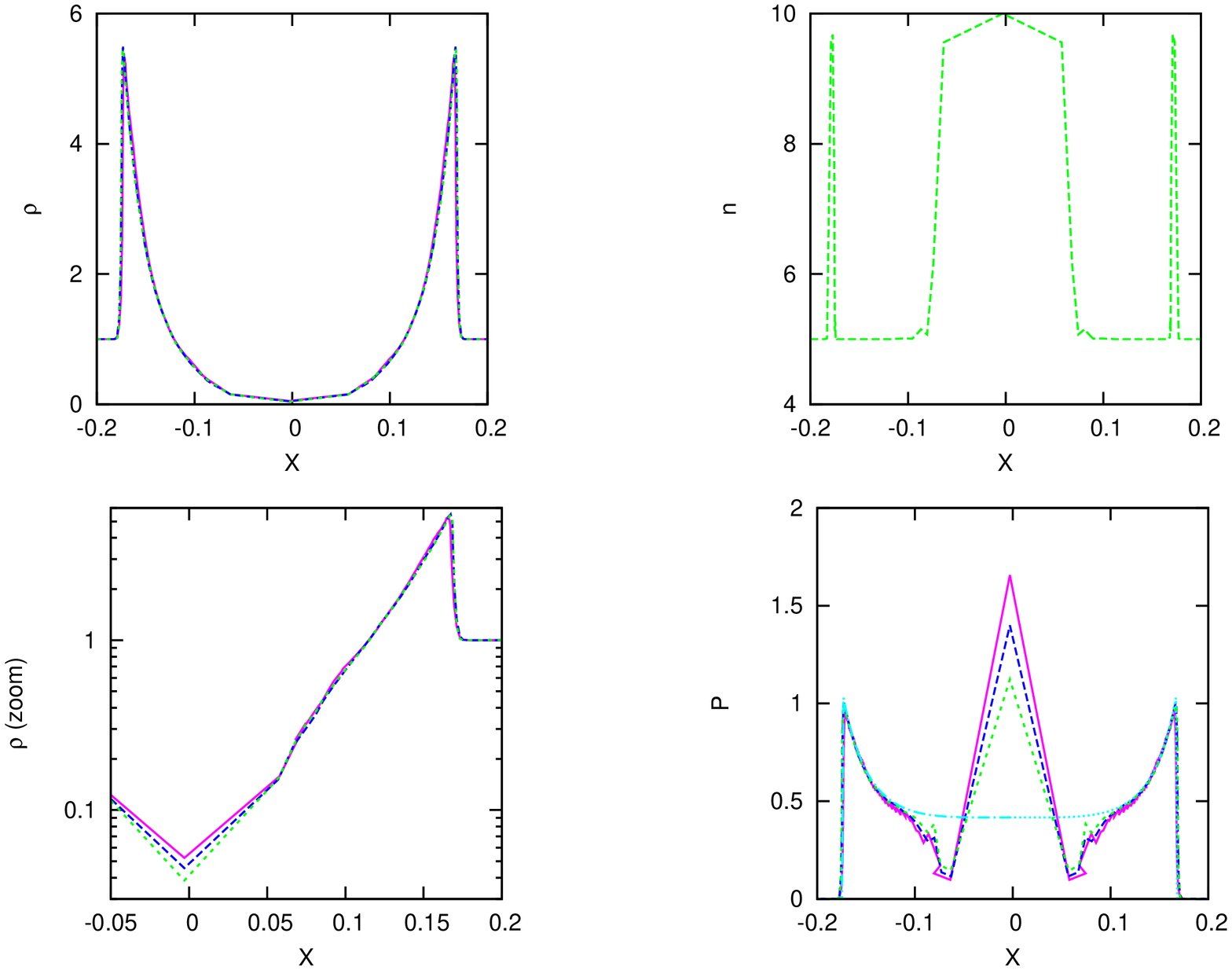}
\begin{figure}
\caption{Density, kernel index, zoom of density and pressure profiles of the 1D shock wave born from a single particle. The
details are the same as in Fig.~\ref{figure5}. }
\label{figure6}
\end{figure}

\clearpage
\includegraphics[angle=-90, width=\textwidth]{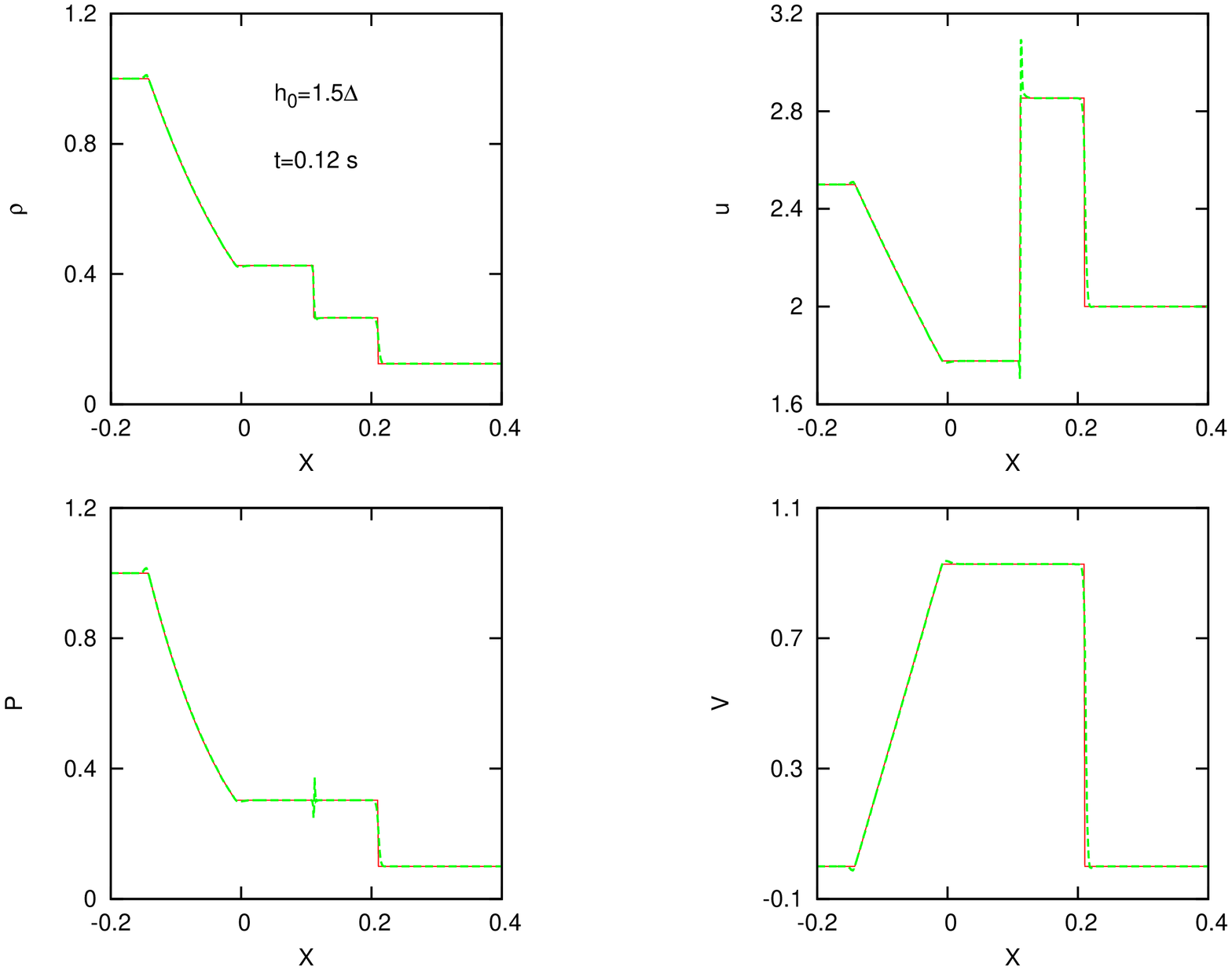}
\begin{figure}
\caption{Characteristic profiles of density, specific internal energy, pressure, and velocity of the 1D shock-tube problem during the self-similar evolution. The continuum red line is the exact solution, the green line was calculated with $n$ adaptive.}  
\label{figure7}
\end{figure}

\clearpage
\includegraphics[angle=-90, width=\textwidth]{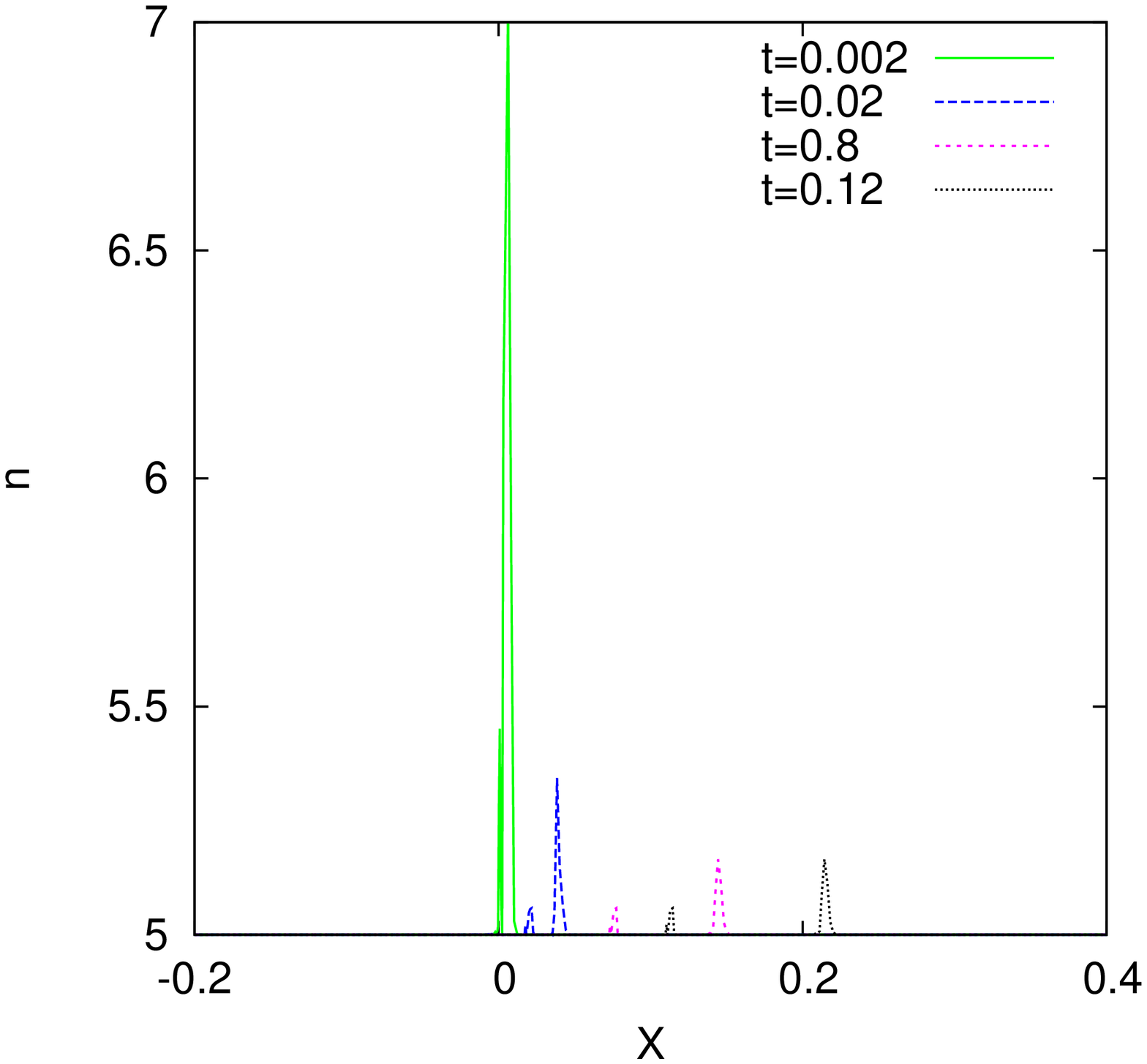}
\begin{figure}
\caption{Profiles of $n$ at different times for the shock-tube numerical experiment.}
\label{figure8}
\end{figure}

\clearpage
\includegraphics[angle=-90, width=\textwidth]{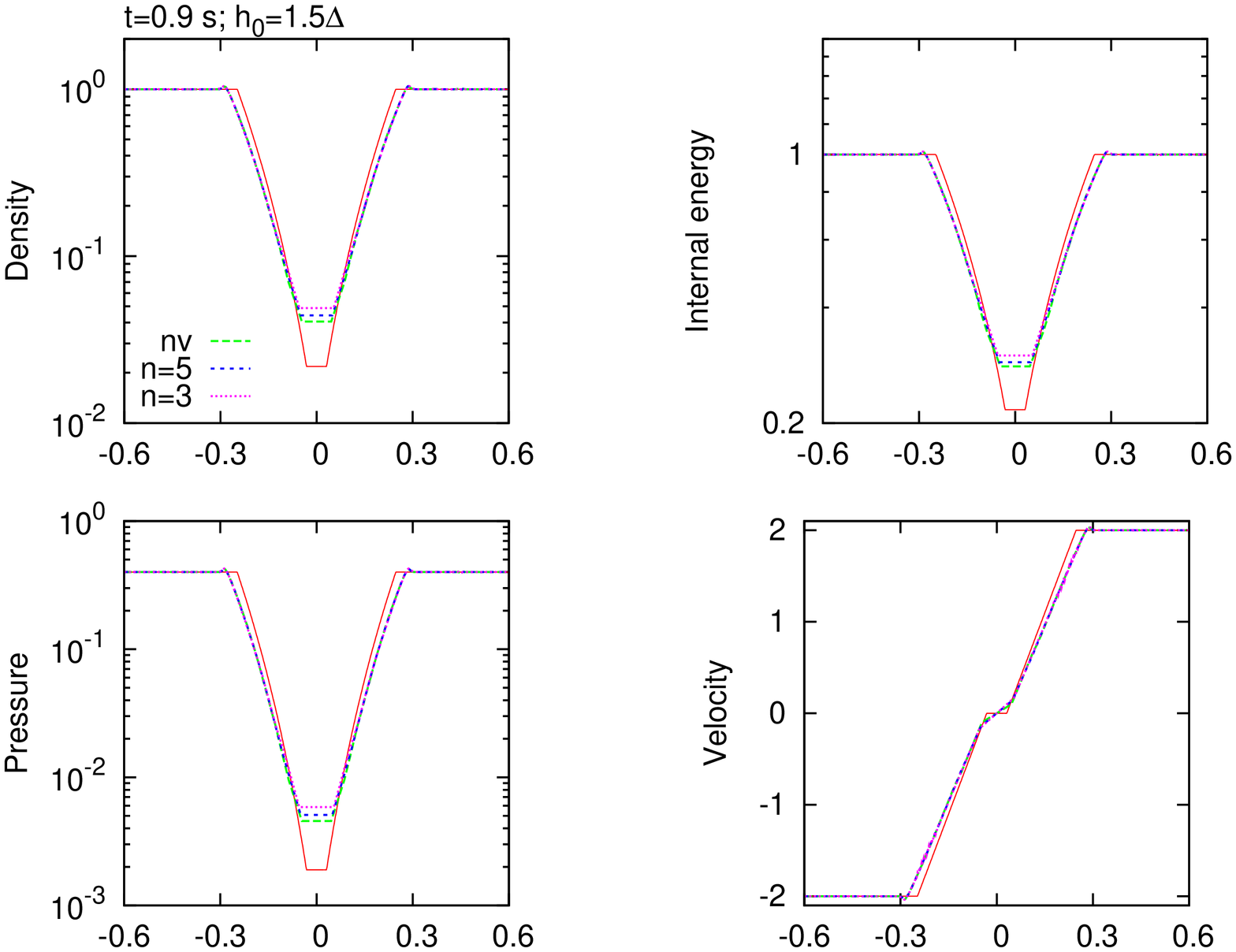}
\begin{figure}
\caption{Density, internal energy, pressure, and velocity profiles for the Sj\"ogreen test with initial particle separation $h_0=1.5\Delta$. The details are the same as in Fig.~\ref{figure5}.}
\label{figure9}
\end{figure}

\clearpage
\includegraphics[angle=-90, width=\textwidth]{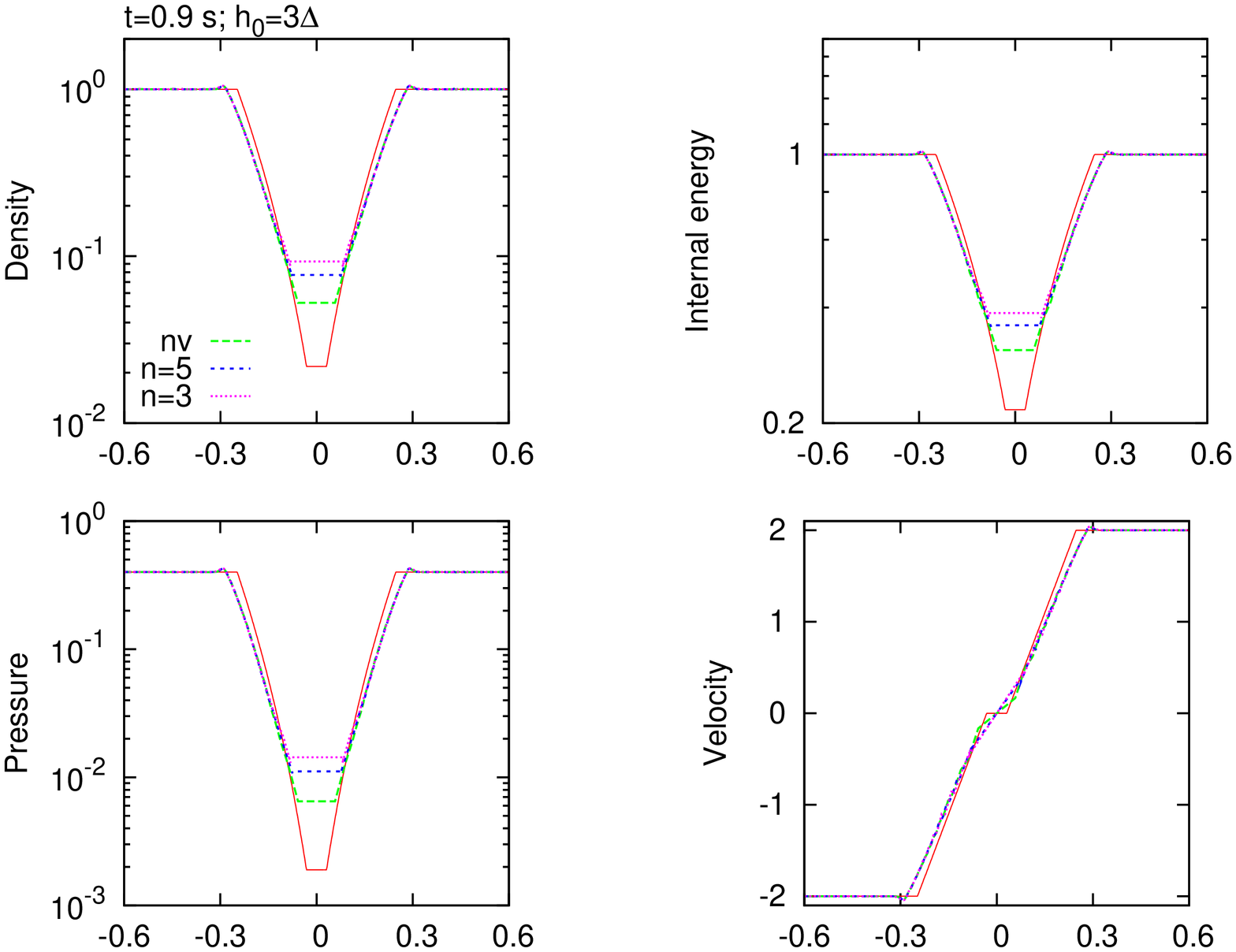}
\begin{figure}
\caption{Same as in Fig.~\ref{figure9}, but for initial particle separation $h_0=3\Delta$.}
\label{figure10}
\end{figure}

\clearpage
\includegraphics[angle=-90, width=\textwidth]{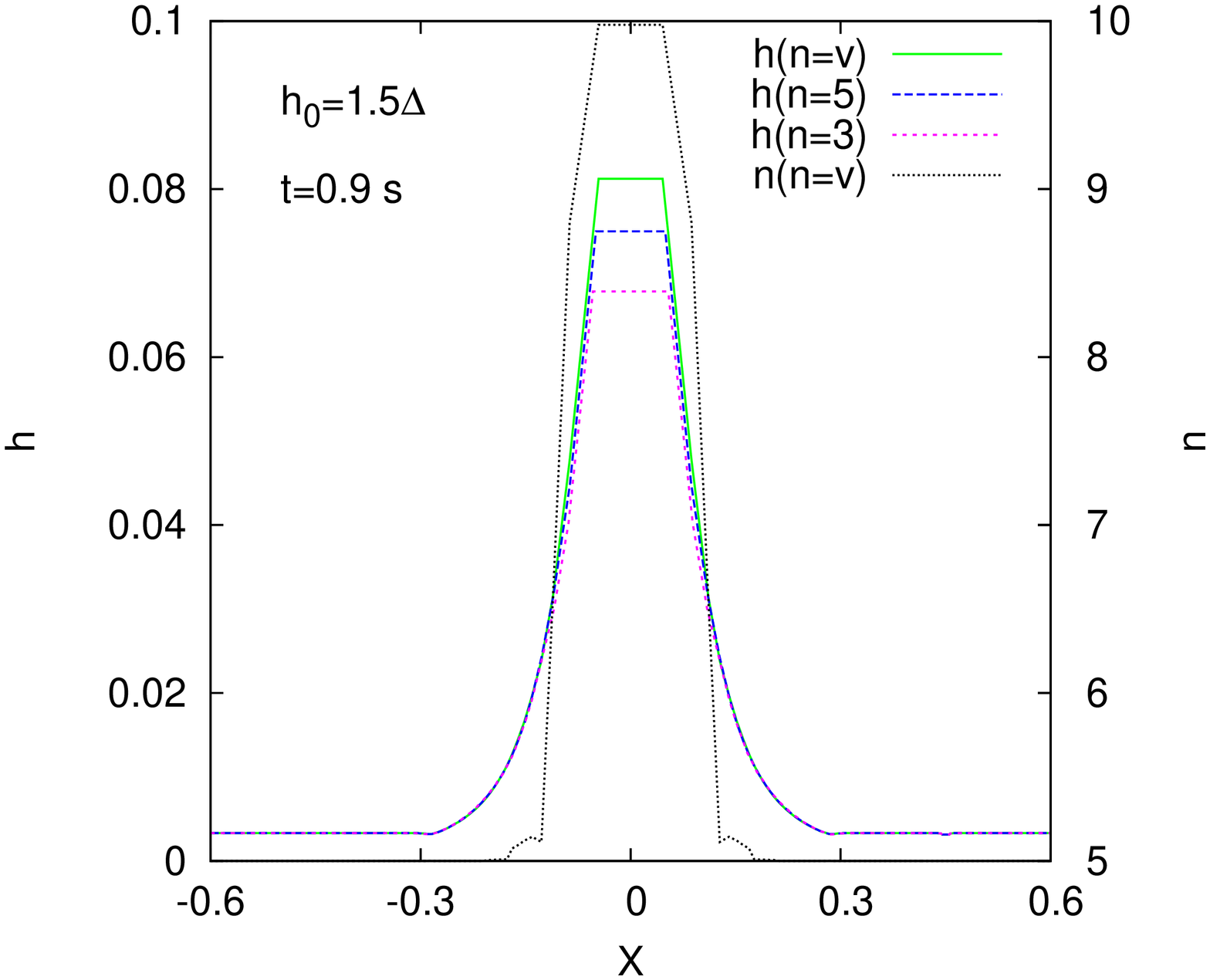}
\begin{figure}
\caption{Profile of the smoothing-length, $h(x)$, for $n=3$ (dashed pink line), $n=5$ (long dashed blue line) and $n$ adaptive (continuum green line) for the Sj\"ogreen test. The profile of $n(x)$ is also shown (black dots). }
\label{figure11}
\end{figure}

\clearpage

\includegraphics[angle=-90, width=\textwidth]{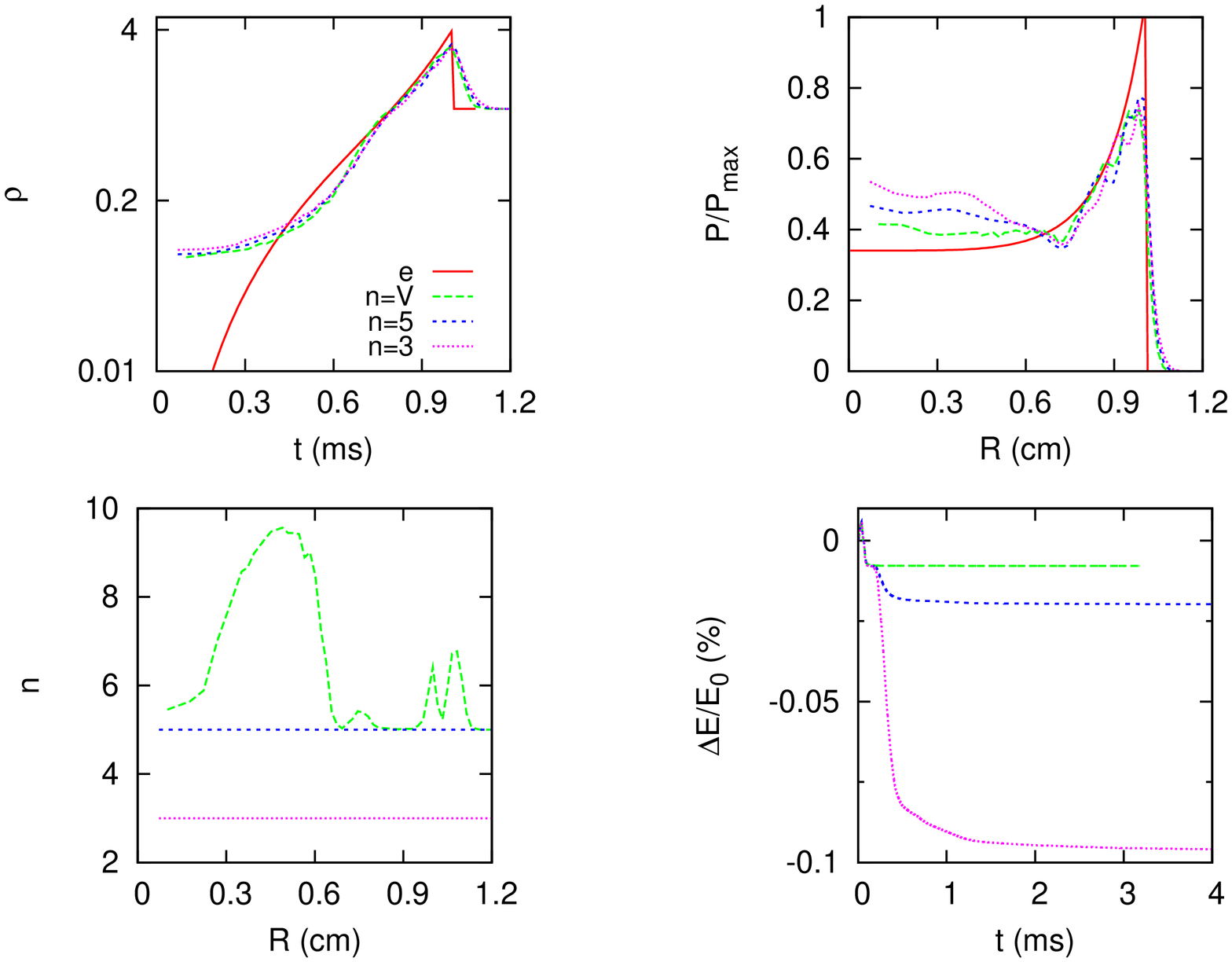}
\begin{figure}
\caption{Profiles of $\rho, P, n$ (averaged in concentric shells) and evolution of energy conservation during the 2D blast wave propagation (Sedov test). The continuum line in red is the classical Sedov solution, and profiles in pink (dots), blue (dashed) and green (long-dashed) are for cases $n=3$, $n=5$ and $n$-adaptive, respectively.} 
\label{figure12}
\end{figure}

\clearpage
\includegraphics[angle=0, width=\textwidth]{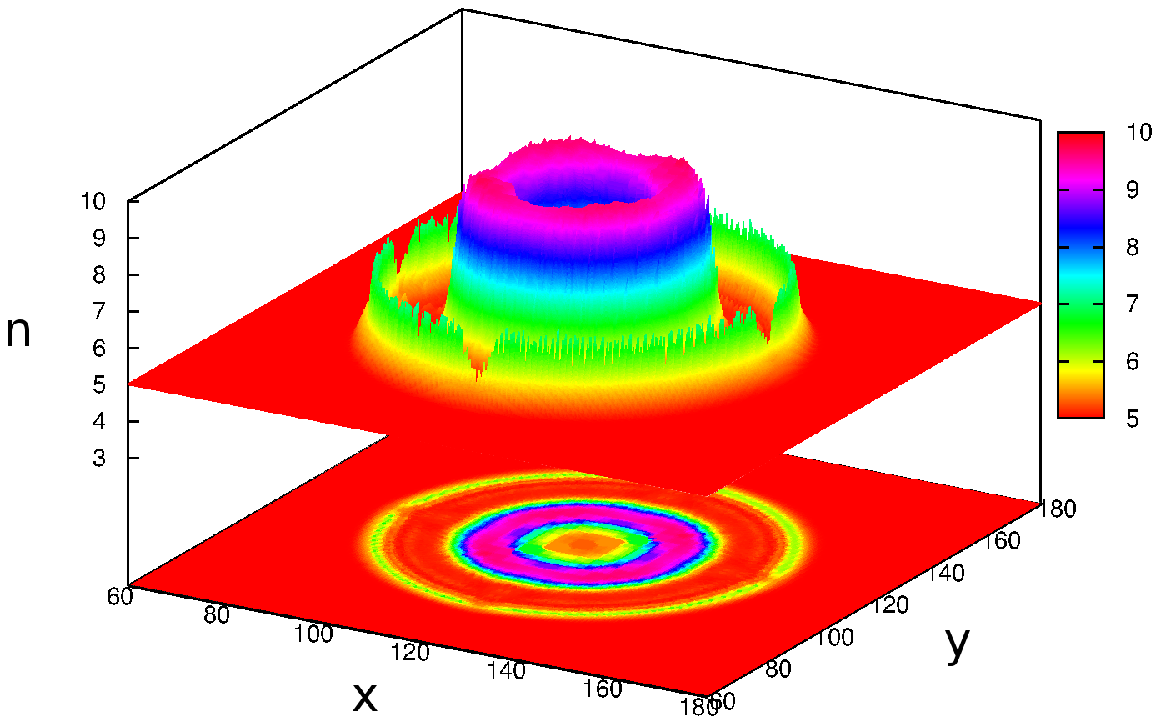}
\begin{figure}
\caption{Rendering of the $sinc$ kernel index $n(x,y)$ for the self-similar wave shown in the bottom left panel of Fig.~\ref{figure12}.} 
\label{figure13}
\end{figure}
\clearpage

\includegraphics[angle=0,  width=\textwidth]{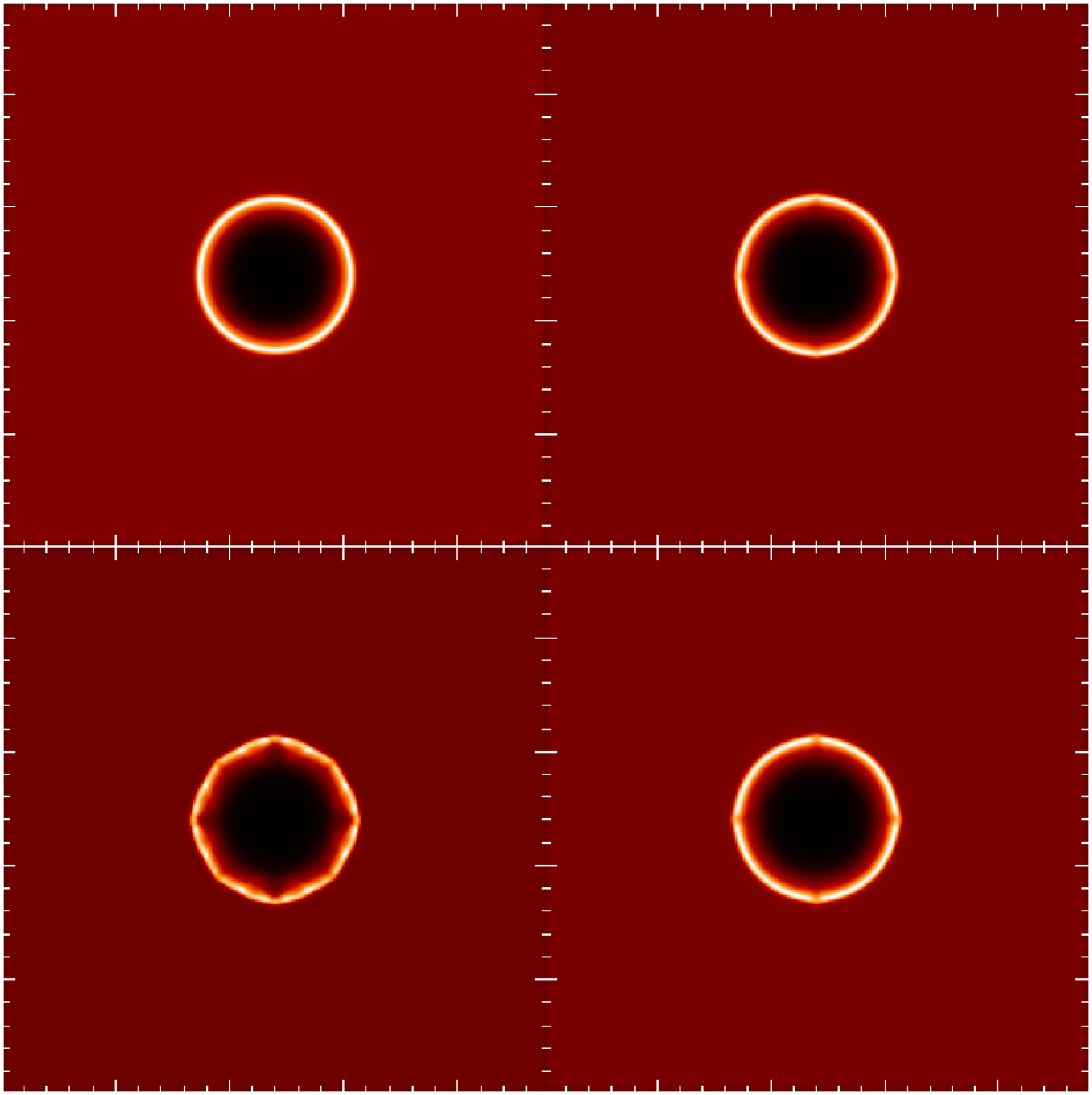}
\begin{figure}
\caption{Density color map of the Sedov wave for $n=3$ (upper
left), $n=5$ (upper right), $n=10$ (bottom left), and $n$-adaptive (bottom right). The spherical symmetry is poorly preserved for a large constant exponent such as $n=10$ in the  $sinc$ kernel.}
\label{figure14}
\end{figure}

\clearpage

\includegraphics[angle=0,  width=\textwidth]{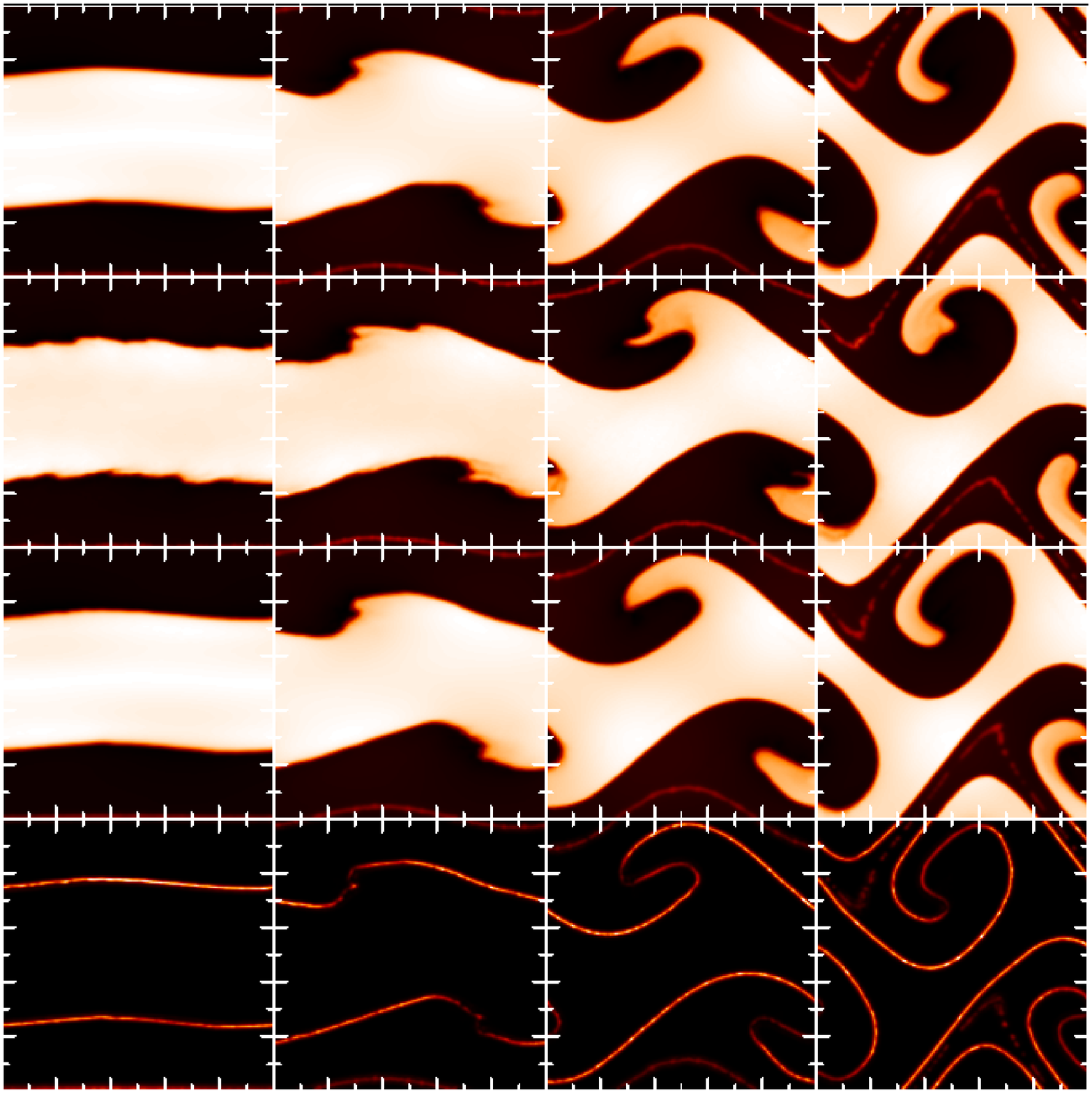}
\begin{figure}
\caption{Density color map showing the growth of the Kelvin-Helmholtz instability for cases $n=5$ (first row), $n=10$ (second row), $n$ adaptive (third row). The last row plots  the contours of $n(x,y)$ for the adaptive case from $n=5$ in the black zones to around $n=7$ in the brightest red zones. } 
\label{figure15}
\end{figure}

\clearpage

\includegraphics[angle=-90,  width=\textwidth]{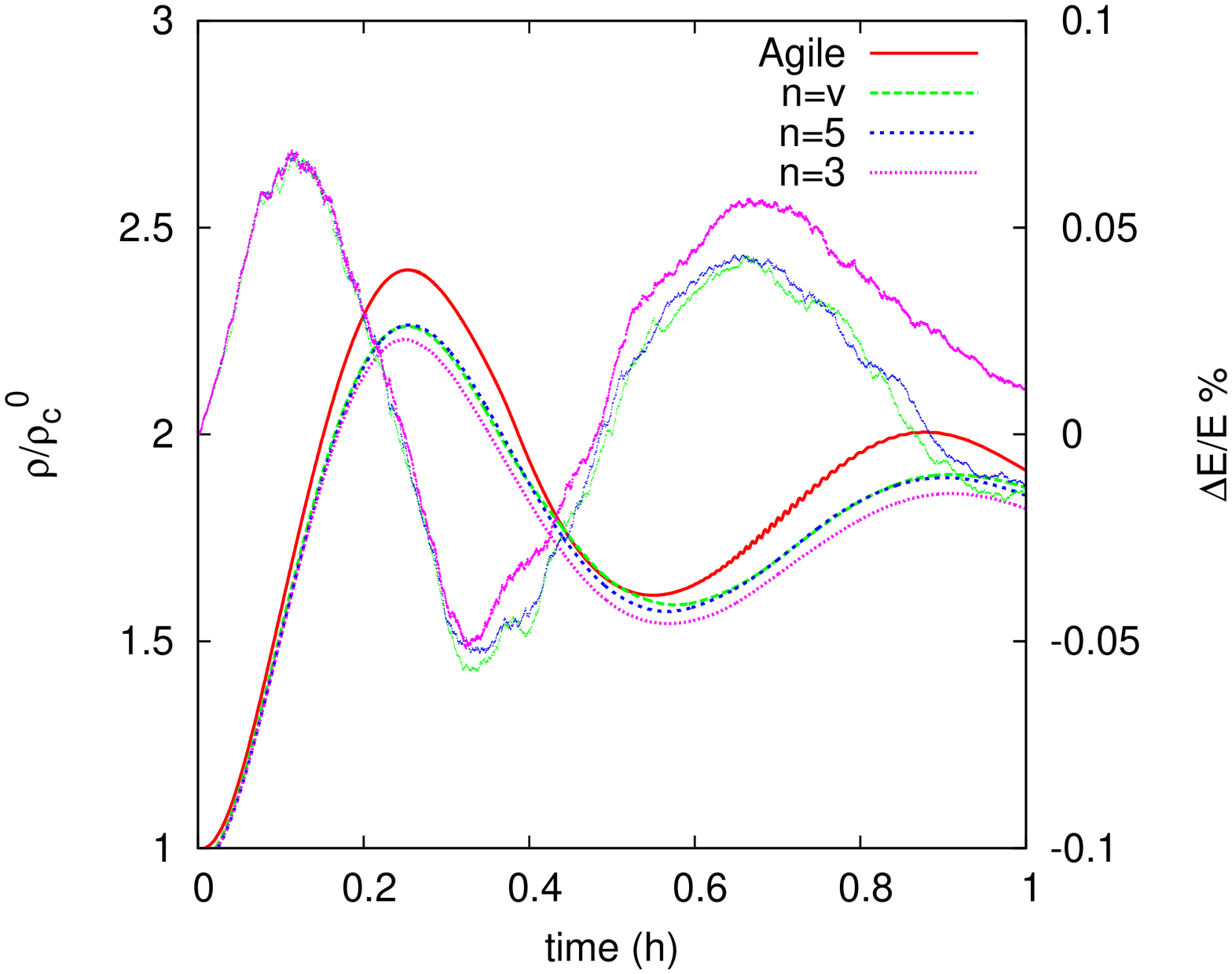}
\begin{figure}
\caption{ Trajectory of the central density during the implosion of a Sun-like polytrope and percent of energy conservation during the first elapsed hour. The red continuum line is the 1D calculation with the implicit Lagrangian hydrocode AGILE. Dashed lines are for $n=3$, $n=5,$ and $n$ adaptive. Light continuum lines in pink, blue and green show the percent of energy conservation. }  
\label{figure16}
\end{figure}

\clearpage

\includegraphics[angle=-90,  width=\textwidth]{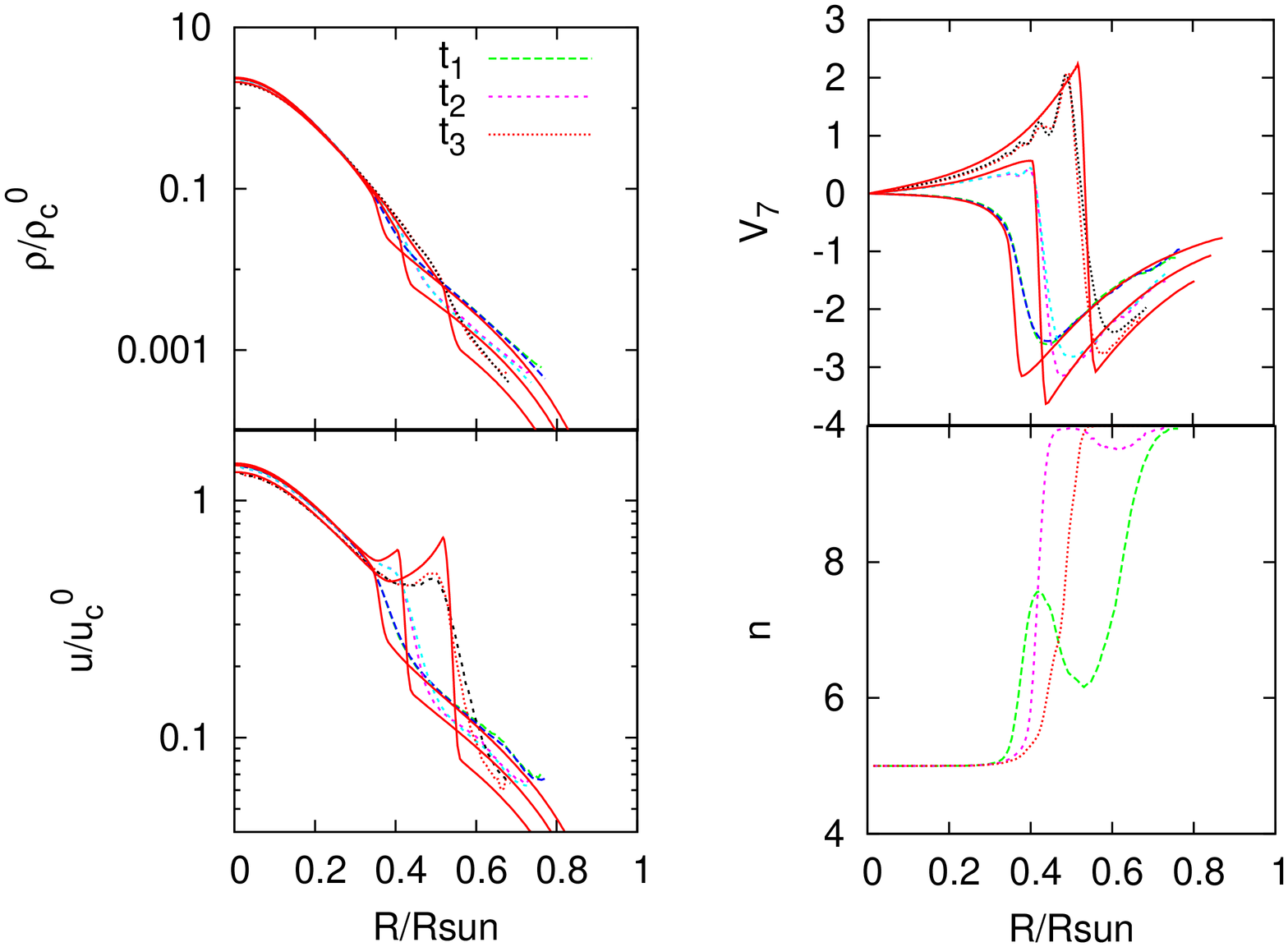}
\begin{figure}
\caption{ Profiles of density, velocity (in units of $10^7$ cm.s$^{-1}$), internal energy, and index of the $sinc$ kernel at times $t_1=870$~s, $t_2=1086$~s, $t_3=1311$~s, corresponding to the collapse of a Sun-like polytrope. The lines in green, pink, and orange are for $n$ adaptive and the blue, light blue, and black  lines for $n=5$ at the same elapsed times. The continuum lines in red have been obtained using a Lagrangian hydrocode with spherical symmetry.}  
\label{figure17}
\end{figure}

\clearpage

\includegraphics[angle=0, scale=0.6]{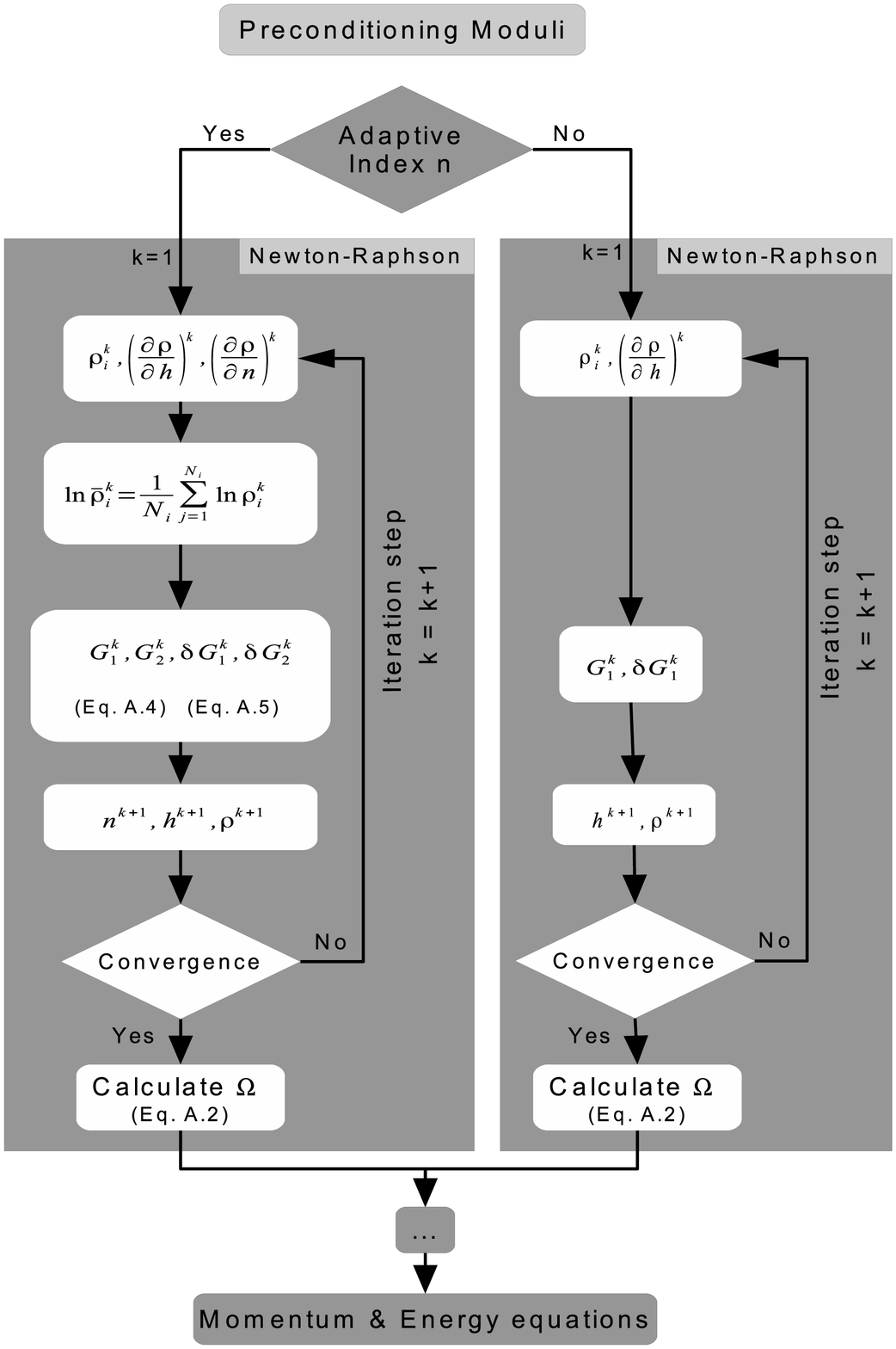}
\begin{figure}
\caption{ Flow chart of the preconditioning algorithm built to set the values of $\rho, h, n$ before starting the hydrodynamics. }  
\label{AppendixFig1}
\end{figure}

\clearpage
\begin{table*}
\centering
\begin{tabular}{@{}llrrrr@{}}
\hline
Dimensionality& $b_0$&$b_1$& $b_2$ & $b_3$  \\
\hline
\hline
1D&$-1.5404568~10^{-2}$&$3.6632876~10^{-1}$ &$-4.6519576~10^{-4}$&$7.3658324~10^{-2}$ \\
2D&$5.2245027~10^{-2}$ &$1.3090245~10^{-1}$&$1.9358485~10^{-2}$ &$-6.1642906~10^{-3}$ \\
3D&$2.7012593~10^{-2}$&$2.0410827~10^{-2}$ &$3.7451957~10^{-3}$&$4.7013839~10^{-2}$ \\
\hline
\end{tabular}
\label{table1}
\caption{Coefficients for calculating the normalization constant $B_n$ in Eq.(\ref{sinc}) }
\end{table*}

\begin{table*}
\centering
\begin{tabular}{@{}llrrrrrrrr@{}}
\hline
Profile&$\rho_0$&$\Delta\rho$& $n_0$ & $\Delta n$ &$\lambda_c$ \\
\hline
\hline
Mountain&0.25&0.75 & 5 &5& 0.5& \\
Valley & 1&0.25 & 5 &5& 0.5& \\
Wall&0.25&0.75 & 5 &5& 0.5&  \\
Cliff&1&- & 5 &5& 0.5&  \\
\hline
\end{tabular}
\label{table1}
\caption{Parameters used in Eqs. (\ref{mountain}) to (\ref{cliff}) (second and third columns) and in Eq.(\ref{indexes}) (last three columns) used to fit several 1D sharp density profiles.}
\end{table*}

\begin{thebibliography}{}





\bibitem[Cabez\'on et al. (2008)]{cabezon08} Cabez\'on R.M., Garc\'\i a-Senz D., Rela\~no A., 2008, J. Comput. Phys., 227, 8523

\bibitem[Cabez\'on et al. (2012)]{cabezon12} Cabez\'on R.M., Garc\'\i a-Senz D., Escart\'in, 2012, A\&A, 545, A112

\bibitem[Dehnen \& Aly (2012)]{dehnen12} Dehnen, W., Aly, H., 2012, MNRAS, 425, 1068

\bibitem[Einfeldt, et al. (1991)]{einfeldt91} Einfeldt, B., Munz, C.D., Roe P.L., Sj\"ogreen, B., 1991, J. Comp. Phys., 92, 273

\bibitem[Garc\'\i a-Senz et al. (2012)]{garciasenz12} Garc\'\i a-Senz D., Cabez\'on R.M., Escart\'in, 2012, A\&A, 538, A9


\bibitem[Gingold \& Monaghan(1977)]{gm77} Gingold R.A., Monaghan J.J, 1977, MNRAS, 181, 375

\bibitem[Hopkins (2012)]{hopkins13}  Hopkins, P.E., 2013, MNRAS, 428, 2840.

\bibitem[Liebend\"orfer et al. (2002)]{liebendorfer02} Liebend\"orfer, M., Rosswog, S., Thielemann, F.K., 2002,  ApJS 141, 229 
\bibitem[Lucy(1977)]{lucy77} Lucy L.B., 1977, AJ, 82, 1013

\bibitem[McNally et al. (2012)]{mcnally12} McNally C., Lyra W., Passy J-C., 2012, ApJ, 201, 18

\bibitem[Monaghan \& Gingold (1983)]{monaghan1983} Monaghan J.J., Gingold, R.A., 1983, J. Comput. Phys.,52,374

\bibitem[Monaghan(1992)]{monaghan92} Monaghan J.J., 1992, ARAA, 365, 199

\bibitem[Monaghan(2005)]{monaghan05} Monaghan J.J., 2005, Rep. Prog. Phys., 68, 1703

\bibitem[Monaghan (1997)]{monaghan97} Monaghan J.J., 1997, J. Comput. Phys.,136, 298

\bibitem[Price (2007)]{price07} Price D., 2007, PASA, 24, 159

\bibitem[Price (2012)]{price2012} Price D., 2012, J. Comput. Phys., 231, 759.


\bibitem[Rosswog (2009)]{rosswog09} Rosswog S., 2009, New Astronomy Review, 53, 78

\bibitem[Rosswog (2014)]{rosswog2014} Rosswog S., 2014, ArXiv:1405.6034R

\bibitem[Sedov (1959)]{sedov1959} Sedov L.I., Similarity and Dimensional Methods in Mechanics. Academic Press Inc. 1959.

\bibitem[Saitoh \& Makino (2013)]{saitoh2013} Saitoh, T, Makino, J., 2013, ApJ, 768, 44 

\bibitem[Sigalotti et al. (2006)]{sigalotti06} L.D.G. Sigalotti, H. L\'opez, A. Donoso, E. Sira, J. Klapp, 2006, J. Comput. Phys., 212, 124-149.


\bibitem[Springel \& Hernquist (2002)]{springel02} Springel, V., Hernquist, L.,  2002, MNRAS , 333, 649.

\bibitem[Springel (2010a)]{springel2010a} Springel, V.  2010a, MNRAS 401, 791.

\bibitem[Springel (2010b)]{springel2010b} Springel, V.  2010b, ARA\&A, 48, 391.


\bibitem[Valdarnini (2012)]{valdarnini12} Valdarnini, R.,  2012, A\&A, 546, A45

\bibitem[Wendland (1995)]{wendland95} Wendland, H.,  1995, Advances in Computational Mathematics, 4, 389

\end{thebibliography}
\end{document}